\pdfoutput=1

\documentclass[journal]{IEEEtran}
\usepackage{cite}

\ifCLASSINFOpdf
\else
\fi
\usepackage[cmex10]{amsmath,mathtools}
\usepackage{amsfonts,amssymb}
\usepackage{amsfonts}

\usepackage{tikz}
\usetikzlibrary{shapes.geometric, arrows}

\tikzstyle{rect} = [rectangle, minimum width=2cm, minimum height=1cm, text centered, draw=black, fill=blue!3]
\tikzstyle{arrow} = [thick,->,>=stealth]

\usepackage{algorithm}
\usepackage{algcompatible} 
\usepackage{algpseudocode}

\usepackage{setspace}

\usepackage[latin1]{inputenc} 

\usepackage{graphicx}
\usepackage{epstopdf}
\usepackage[process=auto]{pstool}
\usepackage{psfrag}

\usepackage{wrapfig}

\usepackage{relsize}

\usepackage[bookmarks=false,implicit=false]{hyperref}

\newcommand{\algorithmfootnote}[2][\footnotesize]{%
  \let\old@algocf@finish\@algocf@finish
  \def\@algocf@finish{\old@algocf@finish
    \leavevmode\rlap{\begin{minipage}{\linewidth}
    #1#2
    \end{minipage}}%
  }%
}

\usepackage{bbm}

\usepackage{bm}

\begin{document}
%
\title{The $\beta$-model for Random Graphs --- Regression, Cram\'er-Rao Bounds, and Hypothesis Testing}
\author{Johan Wahlstr\"{o}m, Isaac Skog,~\IEEEmembership{Member,~IEEE}, Patricio S. La Rosa,~\IEEEmembership{Member,~IEEE}, \\Peter H\"{a}ndel,~\IEEEmembership{Senior Member,~IEEE}, and Arye Nehorai,~\IEEEmembership{Fellow,~IEEE}\vspace*{-5mm}
      \thanks{J. Wahlstr\"{o}m, I. Skog, and P. H\"{a}ndel are with the ACCESS Linnaeus Center, Dept. of Signal Processing, KTH Royal Institute of Technology, Stockholm, Sweden. (e-mail:  \{jwahlst, skog, ph\}@kth.se).

      P. S. La Rosa is with Global IT Analytics, Monsanto Company, and with the School of Engineering, Washington University in St. Louis, MO 63130 USA (e-mail: pslaro@monsanto.com).


      A. Nehorai is with the Preston M. Green Department of Electrical and Systems Engineering, Washington University in St. Louis, MO 63130 USA (e-mail: nehorai@ese.wustl.edu).

      The work of A. Nehorai was supported by AFOSR Grant No. FA9550-11-1-0210.
      }
}


%


\maketitle
\begin{abstract}
We develop a maximum-likelihood based method for regression in a setting where the dependent variable is a random graph and covariates are available on a graph-level. The model generalizes the well-known $\beta$-model for random graphs by replacing the constant model parameters with regression functions. Cram\'er-Rao bounds are derived for the undirected $\beta$-model, the directed $\beta$-model, and the generalized $\beta$-model. The corresponding maximum likelihood estimators are compared to the bounds by means of simulations. Moreover, examples are given on how to use the presented maximum likelihood estimators to test for directionality and significance. Last, the applicability of the model is demonstrated using dynamic social network data describing communication among healthcare workers.
\end{abstract}



%
\IEEEpeerreviewmaketitle

\section{Introduction}

%
%
%
%
%
%
%
%
%
%
%

A random graph is a random variable whose realizations are graphs. Typically, the number of nodes is considered to be fixed. Random graphs have been used to model e.g.,
interactome networks, with nodes and edges representing molecules and their biological interaction, respectively \cite{Vidal2011}; brain networks, with nodes and edges representing brain regions and their structural or functional connectivity, respectively \cite{LaRosa2016,Andersen2012}; and social networks, with nodes and edges representing social actors (such as individuals or organizations) and their interaction, respectively \cite{Robins2007}. In many of these applications, variations in the observed graphs correlate with variations in external covariates. For example, properties of the interactome relate to human disease and the function of specific proteins \cite{Sharan2007}, brain connectivity is closely related to demographic and psychometric measures \cite{Smith2015}, and social interaction is related to both external events and the demographics and spatial distribution of the studied population \cite{Dong2014}. Despite numerous examples of phenomena that can be modeled as
dynamic or temporal networks \cite{Holme2012}, there is still a need to extend or generalize existing statistical methods for static graphs to enable the incorporation of dynamic side information.

The aim of the present study is to develop methods for point estimation, uncertainty estimation, and hypothesis testing, in a setting where the considered observations are graphs, and where covariate information is available on a graph-level. This is done by starting from an already established model for random graphs, namely the $\beta$-model, and then generalizing it so as to enable multivariate regression analysis. In the remainder of this section, we describe the considered graph models, and review previously employed methods for incorporating covariate information into random graphs.

The $\beta$-model belongs to the class of exponential random graphs models (ERGMs), also known as $p^*$ models, which is the subset of the exponential family that describes random graphs \cite{Holland1981,Frank1986,Strauss1986,Newman2003,Schmidt2013}. Among the ERGMs studied in the literature, the sufficient statistics defining a specific ERGM have included reciprocity measures (describing the tendency of mutual connection between two nodes in a directed graph) \cite{Holland1981}, the number of k-stars (formations with one node connected to $k$ other nodes but with no additional edges between these $k$ nodes), and the number of triangles (three mutually connected nodes) \cite{Frank1986}. The main appeal of ERGMs is that their probability distribution can be specified in terms of any graph attribute that may have relevance for the modeled network.

One of the most simple ERGMs is obtained by letting the sufficient statistics be the vector specifying the degree of each node in the graph \cite{Rinaldo2013}. This is a special case of the model for logistic regression, known as the $\beta$-model \cite{Skyler2011,Chatterjee2011}. The popularity of the $\beta$-model can be attributed to its flexibility in adjusting to observed degree sequences. Specifically, in a $\beta$-model where the model parameters are obtained by using the method of maximum likelihood (ML), the degree of each node has an expectation value equal to the observed degree of the same node \cite{Chatterjee2011}. Since the ML estimates of the parameters in the $\beta$-model cannot be written in closed form, several iterative or Monte Carlo-based estimation procedures have been proposed \cite{Snijders02}. These have included iterative scaling, Newton's method, Fisher's method of scoring, iteratively reweighted least squares \cite{Holland1981}, \cite{Minka2003}, and other
fixed-point iteration schemes \cite{Chatterjee2011}.

Information from exogenous variables are often incorporated into random graphs by using nodal (associated with a node) or dyadic (associated with a pair of nodes) covariates. For example, in a social network, a nodal covariate could represent gender or status, whereas a dyadic covariate could represent absolute age difference or spatial distance. Random graphs with nodal covariates are commonly modeled as stochastic block models (SBMs) \cite{Fienberg1981,Holland1983,Wang1987,Celisse2012}. In a SBM, the nodes are divided into groups (blocks), with each group collecting all nodes with some given covariate values. The distribution of the random variable describing the relation between two nodes is then constrained to only depend on the groups to which the two nodes belong. ERGMs with nodal and dyadic covariates have previously been studied in \cite{Suesse2012} and \cite{VanDuijn2004}.

In this article, we generalize the $\beta$-model by incorporating covariates on a graph-level. This allows us to perform regression with a random graph as the dependent variable. ML estimates are obtained by generalizing the fixed-point iteration scheme that was proposed for the $\beta$-model in \cite{Chatterjee2011}. To characterize the considered estimation problems, Cram\'er-Rao bounds (CRBs) are derived for the undirected, directed, and generalized $\beta$-model, and the performance of the ML estimator is illustrated by means of simulations. Further, examples are given on how to use the presented ML estimators to perform significance tests and tests for directionality. The applicability of the estimation framework is demonstrated in a case study where we use real-world data describing the interaction among ten healthcare workers. The resulting social network is regressed on categorical covariates representing the time of day and the day of the week.

%
%
%
%
%
%
%
%



\begin{figure}[t]
\def\svgwidth{0.5\columnwidth}
\hspace*{5mm}
\scalebox{1.5}{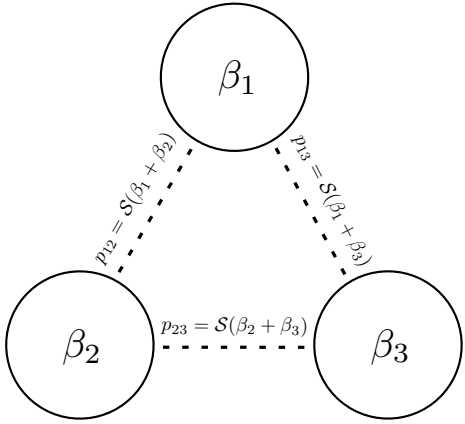}
\vspace*{-1mm}
\caption{Illustration of the undirected $\beta$-model with $n=3$ nodes.}
\label{undirectedFig}
\vspace*{-1mm}
\end{figure}

\section{Estimation framework}
\label{FrameworkSection}

To lay the foundation for the generalized $\beta$-model and the results in later sections, we review the undirected and directed $\beta$-models in Section \ref{undirbetasec} and Section \ref{directedmodelsubsection}, respectively. The generalized $\beta$-model is then introduced in Section \ref{generalizedmodelsubsection}. The notation is defined independently in each subsection. All graphs are assumed to be unweighted and without self-loops.

\subsection{The Undirected $\beta$-model}
\label{undirbetasec}

Consider a graph with $n$ nodes. In the undirected $\beta$-model,
the probability of having an edge between the $i$th node and the $j$th node is
\begin{equation}
\label{edgeprobabilityundirected}
p_{ij}=\frac{e^{\beta_i+\beta_j}}{1+e^{\beta_i+\beta_j}}
\end{equation}
for any $i\neq j$. Hence, the $n(n-1)/2$ edge probabilities are parameterized by the $n$ parameters $\beta_1,\dots,\beta_n$. As may be realized from \eqref{edgeprobabilityundirected}, $\beta_i$ signifies the tendency of the $i$th node to form edges with other nodes, i.e., its so called \emph{differential attractiveness} \cite{Holland1981}. Specifically, the $i$th node can be expected to have a large (small) number of ties if $\beta_i$ is positive (negative) and of large magnitude. For ease of notation, we set $p_{ii}=0$ for $i=1,\dots,n$. The model is illustrated in Fig. \ref{undirectedFig}, where ${\cal S}(\beta)\overset{_\Delta}{=}1/(1+e^{-\beta})$ denotes the sigmoid function.

Now, let $Y_{ij}=Y_{ji}$ denote the number of times that an edge between node $i$ and node $j$ is present in $N_{ij}=N_{ji}$ independent measurements. It then follows that
\begin{equation}
\label{measdistundirectedeq}
Y_{ij}\sim\mathrm{Bin}(N_{ij},p_{ij})
\end{equation}
where $\mathrm{Bin}(\cdot\,,\cdot)$ denotes the binomial distribution with the first and second parameter indicating the number of observations and the probability of success, respectively. Here, we let $N_{ii}=0$ for $i=1,\dots,n$. Moreover, the likelihood function of $\boldsymbol{\theta}\overset{_\Delta}{=}[\hspace*{0.2mm}\beta_1\;\dots\;\beta_n\hspace*{0.2mm}]^{\intercal}$, conditioned on $N_{ij}$ observations associated with each unordered pair of nodes $(i,j)$, is proportional to \cite{Chatterjee2011}
\begin{equation}
\label{likundirectedeq}
L(\boldsymbol{\theta}|\mathbf{Y})=\frac{e^{\sum_{i=1}^{n}\beta_id_i}}{\prod_{i=1,\hspace*{0.3mm}j=i+1}^n(1+e^{\beta_i+\beta_j})^{N_{ij}}}.
\end{equation}
Here, $\mathbf{Y}\overset{_\Delta}{=}\{Y_{ij}\}_{i=1,\hspace*{0.3mm}j=i+1}^n$, while the degree of node $i$, summed over all observations, have been defined as $d_i\overset{_\Delta}{=}\sum_{j=1}^{n}Y_{ij}$. We will, with some abuse of terminology, continue to refer to $d_i$ as a degree. Since the Hessian of $L(\boldsymbol{\theta}|\mathbf{Y})$ with respect to $\boldsymbol{\theta}$ can easily be shown to be negative semidefinite, the ML estimates $\hat{\boldsymbol{\theta}}\overset{_\Delta}{=}[\hspace*{0.2mm}\hat{\beta}_1\;\dots\;\hat{\beta}_n\hspace*{0.2mm}]^{\intercal}$ must necessarily satisfy
$\partial L(\hat{\boldsymbol{\theta}}|\mathbf{Y})/\partial\beta_i=0$ for $i=1,\dots,n$. This is equivalent to saying that \cite{Chatterjee2011}
\begin{equation}
\label{derivequndirected}
d_i=\sum_{j=1}^{n}N_{ij}\frac{e^{\hat{\beta}_i+\hat{\beta}_j}}{1+e^{\hat{\beta}_i+\hat{\beta}_j}}
\end{equation}
for $i=1,\dots,n$. The interpretation of \eqref{derivequndirected} is that the observed degree of each node must be equal to the expectation value of the corresponding degree under the model that is implied by the ML estimates. This interpretation makes use of the fact that expected value of $Y_{ij}$ is $N_{ij}p_{ij}$. To find the ML estimates, \cite{Chatterjee2011} introduced the function $\boldsymbol{\varphi}:\mathbb{R}^n\rightarrow\mathbb{R}^n$, whose outputs are defined according to
\begin{equation}
\label{phifncundirected}
\varphi_i(\mathbf{z})\overset{_\Delta}{=}\log(d_i)-\log \sum_{j=1}^{n}\frac{N_{ij}}{e^{-z_j}+e^{z_i}}
\end{equation}
for $i=1,\dots,n$, and where we have used that $\mathbf{z}\overset{_\Delta}{=}[\hspace*{0.2mm}z_1\;\dots\;z_n\hspace*{0.2mm}]^{\intercal}$. Rearranging the terms in \eqref{phifncundirected} and comparing with \eqref{derivequndirected} it can be seen that
$\hat{\boldsymbol{\theta}}$ is a fixed point of $\boldsymbol{\varphi}$. Hence, the ML estimates may be found by iterating
\begin{equation}
\mathbf{z}^{(m+1)}=\boldsymbol{\varphi}(\mathbf{z}^{(m)})
\end{equation}
until convergence, starting from some initial value $\mathbf{z}^{(0)}$. Assuming that a unique ML solution exists and that $N_{ij}=1$ for all $i\neq j$, geometrically fast convergence to the fixed point was shown in \cite{Chatterjee2011} and \cite{Csiszar2011}. Conditions for the asymptotic normality of the ML estimator when $n$ tends to infinity and $N_{ij}=1$ for all $i\neq j$ were presented in \cite{Yan2013}. Necessary and sufficient conditions for the existence of a finite ML estimate were presented in \cite{Rinaldo2013}, \cite{Csiszar2011}, and \cite{Karwa2016}. As an example, it should be clear that there is no finite $\hat{\boldsymbol{\theta}}$ satisfying \eqref{derivequndirected} whenever $d_i=\textstyle\sum_{j=1}^{n}N_{ij}$ or $d_i=0$ for some $i\in\{1,\dots,n\}$.

\begin{figure}[t]
\def\svgwidth{0.5\columnwidth}
\hspace*{5mm}
\scalebox{1.5}{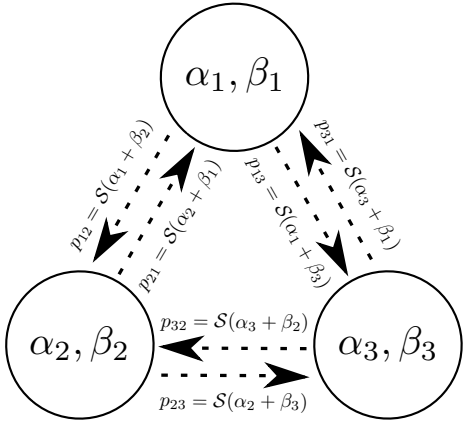}
\vspace*{-1mm}
\caption{Illustration of the directed $\beta$-model with $n=3$ nodes.}
\label{directedFig}
\vspace*{-1mm}
\end{figure}

\subsection{The Directed $\beta$-model}
\label{directedmodelsubsection}

In the directed $\beta$-model \cite{Holland1981}, the probability of having an edge directed from the $i$th node to the $j$th node is
\begin{equation}
\label{edgeprobabilitydirected}
p_{ij}=\frac{e^{\alpha_i+\beta_j}}{1+e^{\alpha_i+\beta_j}}
\end{equation}
for any $i\neq j$, while $p_{ii}=0$ for $i=1,\dots,n$. Assuming $n$ nodes, this means that the $n(n-1)$ edge probabilities are parameterized by the $2n$ parameters 
$\alpha_1,\dots,\alpha_n,\beta_1,\dots,\beta_n$. As should be obvious, $\alpha_i$ affects the probability of having an edge directed \emph{from} the $i$th node, while $\beta_i$ affects the probability of having an edge directed \emph{to} the $i$th node. The parameters $\alpha_i$ and $\beta_i$ are called \emph{productivity} and \emph{attractiveness} parameters, respectively \cite{Holland1981}. To resolve the additive ambiguity of the parameters ($p_{ij}$ is not altered when adding a constant to $\alpha_i$ for $i=1,\dots,n$ and simultaneously subtracting the same constant from $\beta_i$ for $i=1,\dots,n$), we will without loss of generality assume that $\beta_n$ is known to be zero. The model is illustrated in Fig. \ref{directedFig}.

Now, let $Y_{ij}$ denote the number of times that an edge directed from node $i$ to node $j$ is present in $N_{ij}$ independent measurements. Once again, we let $N_{ii}=0$ for $i=1,\dots,n$. It then follows that
\begin{equation}
\label{measdistdirectedeq}
Y_{ij}\sim \mathrm{Bin}(N_{ij},p_{ij}).
\end{equation}
Further, assuming that we have $N_{ij}$ observations associated with each ordered pair of nodes $(i,j)$, the likelihood function of $\boldsymbol{\theta}\overset{_\Delta}{=}[\hspace*{0.2mm}\alpha_1\;\dots\;\alpha_n\;\beta_1\;\dots\;\beta_{n-1}\hspace*{0.2mm}]\rule{0pt}{6.7pt}^{\intercal}$ is proportional to
\begin{equation}
\label{likelihoodfncdirectedeq}
L(\boldsymbol{\theta}|\mathbf{Y}) = \frac{e^{\sum_{i=1}^{n}\alpha_i^{}d_{i}^{\alpha}+\beta_i^{}d_{i}^{\beta}}}{\prod_{i,j=1}^n (1+e^{\alpha_i+\beta_j})^{N_{ij}}}.
\end{equation}
Here, $\mathbf{Y}\overset{_\Delta}{=}\{Y_{ij}\}_{i,j=1}^n$, while the outdegree and indegree of node $i$, summed over all observations, have been defined as $d_{i}^{\alpha}\overset{_\Delta}{=}\textstyle\sum_{j=1}^{n}Y_{ij}$ and $d_{i}^{\beta}\overset{_\Delta}{=}\textstyle\sum_{j=1}^{n}Y_{ji}$, respectively. The ML estimates $\hat{\boldsymbol{\theta}}\overset{_\Delta}{=}[\hspace*{0.2mm}\hat{\alpha}_1\;\dots\;\hat{\alpha}_n\;\hat{\beta}_1\;\dots\;\hat{\beta}_{n-1}\hspace*{0.2mm}]\rule{0pt}{6.7pt}^{\intercal}$ must necessarily satisfy
$\partial L(\hat{\boldsymbol{\theta}}|\mathbf{Y})/\partial\alpha_i=0$ for $i=1,\dots,n$ and $\partial L(\hat{\boldsymbol{\theta}}|\mathbf{Y})/\partial\beta_i=0$ for $i=1,\dots,n-1$. This is equivalent to saying that
\begin{subequations}
\label{deriveq}
\begin{align}
d_i^{\alpha}&=\sum_{j=1}^{n}N_{ij}\frac{e^{\hat{\alpha}_i+\hat{\beta}_j}}{1+e^{\hat{\alpha}_i+\hat{\beta}_j}},
\intertext{for $i=1,\dots,n$, where we have used that $\hat{\beta}_n=0$, and}
d_i^{\beta}&=\sum_{j=1}^{n}N_{ji}\frac{e^{\hat{\alpha}_j+\hat{\beta}_i}}{1+e^{\hat{\alpha}_j+\hat{\beta}_i}},
\end{align}
\end{subequations}
for $i=1,\dots,n-1$. In similarity with \eqref{derivequndirected}, equation \eqref{deriveq} can be interpreted as saying that the observed outdegree and indegree of each node must be equal to the expectation values of the corresponding degrees under the model that is implied by the ML estimates. To find the ML estimates, we introduce the function $\boldsymbol{\varphi}:\mathbb{R}^{2n-1}\rightarrow\mathbb{R}^{2n-1}$ whose outputs are defined according to
\begin{subequations}
\label{phieq}
\begin{equation}
\varphi_i(\mathbf{z})\overset{_\Delta}{=}\log(d_i^{\alpha})-\log\sum_{j=1}^{n}\frac{N_{ij}}{e^{-z_{j+n}}+e^{z_i}},
\end{equation}
for $i=1,\dots,n$, and
\begin{equation}
\varphi_{i+n}(\mathbf{z})\overset{_\Delta}{=}\log(d_i^{\beta})-\log \sum_{j=1}^{n}\frac{N_{ji}}{e^{-z_j}+e^{z_{i+n}}},
\end{equation}
\end{subequations}
for $i=1,\dots,n-1$. Here, we have used that $\mathbf{z}\overset{_\Delta}{=}[\hspace*{0.2mm}z_1\;\dots\;z_{2n-1}\hspace*{0.2mm}]^{\intercal}$ and $z_{2n}\overset{_\Delta}{=}0$. Rearranging the terms in \eqref{phieq} and comparing with \eqref{deriveq} it can be seen that 
$\hat{\boldsymbol{\theta}}$
is a fixed point of $\boldsymbol{\varphi}(\mathbf{z})$. Hence, we may, just as for the undirected model, find the ML estimates by iterating
\begin{equation}
\mathbf{z}^{(m+1)}=\boldsymbol{\varphi}(\mathbf{z}^{(m)})
\end{equation}
until convergence, starting from some initial value $\mathbf{z}^{(0)}$.

\begin{figure}[t]
\def\svgwidth{0.5\columnwidth}
\hspace*{5mm}
\scalebox{1.5}{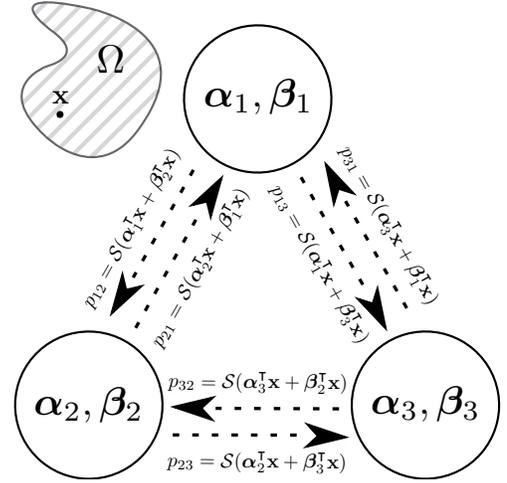}
\vspace*{-1mm}
\caption{Illustration of the generalized $\beta$-model with $n=3$ nodes.}
\label{generalizedFig}
\vspace*{-1mm}
\end{figure}

\subsection{The Generalized $\beta$-model}
\label{generalizedmodelsubsection}

We will now demonstrate how to generalize the $\beta$-model for the purpose of performing regression. Using this generalization, the estimated model parameters will not only provide us with a static description of the random graph, but also reveal how the random graph varies with the chosen covariates. The core idea is to replace the parameters $\alpha_i$ and $\beta_j$ with the regression functions $\boldsymbol{\alpha}_i^{\intercal}\mathbf{x}$ and
$\boldsymbol{\beta}_j^{\intercal}\mathbf{x}$, respectively. Hence, the probability of having an edge directed from the $i$th node to the $j$th node becomes
\begin{equation}
\label{edgeprobabilitygeneralized}
p_{ij}(\mathbf{x})=\frac{e^{\boldsymbol{\alpha}_i^{\intercal}\mathbf{x}+\boldsymbol{\beta}_j^{\intercal}\mathbf{x}}}{1+e^{\boldsymbol{\alpha}_i^{\intercal}\mathbf{x}+\boldsymbol{\beta}_j^{\intercal}\mathbf{x}}}
\end{equation}
for any $i\neq j$, while $p_{ii}(\mathbf{x})=0$ for $i=1,\dots,n$.
The probability depends on both the $2K$ regression coefficients $\boldsymbol{\alpha}_i\overset{_\Delta}{=}[\hspace*{0.2mm}\alpha_{i,1}\;\dots\;\alpha_{i,K}\hspace*{0.2mm}]^{\intercal}$ and $\boldsymbol{\beta}_j\overset{_\Delta}{=}[\hspace*{0.2mm}\beta_{j,1}\;\dots\;\beta_{j,K}\hspace*{0.2mm}]^{\intercal}$, and the $K$ covariates $\mathbf{x}=[\hspace*{0.2mm}x_1^{}\;\dots\;x_K^{}\hspace*{0.2mm}]^{\intercal}\in\Omega$. Here, $\alpha_{i,k}$ and $\beta_{i,k}$ describe the effect that the $k$th covariate has on the tendency of the $i$th node to form edges with other nodes. For example, if $\alpha_{i,k}$ is positive (negative), the probability of having an edge directed from the $i$th node increases (decreases) as $x_k$ increases. In analogy with the preceding subsection, we let $\boldsymbol{\beta}_{n}=\mathbf{0}_{K\hspace*{-0.2mm},1}$ where $\mathbf{0}_{\ell_1\hspace*{-0.2mm},\,\ell_2}$ is the zero matrix of dimension $\ell_1\times\ell_2$. The model is illustrated in Fig. \ref{generalizedFig}.


Now, assume that we know the covariate values $\{\mathbf{x}_{\ell}\}_{\ell=1}^L$ and have made the associated observations $\mathbf{Y}\overset{_\Delta}{=}\{\{Y_{ij\hspace*{-0.2mm},\ell}\}_{i,j=1}^n\}_{\ell=1}^L$ from a set of $L$ random graphs, each with $n$ nodes. Here, $Y_{ij\hspace*{-0.2mm},\ell}$ is the number of times that an edge directed from node $i$ to node $j$ is present in $N_{ij\hspace*{-0.2mm},\ell}$ independent measurements from random graph $\ell$, where $N_{ii\hspace*{-0.2mm},\ell}=0$ for $i=1,\dots,n$ and $\ell=1,\dots,L$. It then holds that
\begin{equation}
\label{measdistgeneralizedeq}
Y_{ij\hspace*{-0.2mm},\ell}\sim\mathrm{Bin}(N_{ij\hspace*{-0.2mm},\ell},p_{ij}(\mathbf{x}_{\ell})).
\end{equation}
Furthermore, it can be seen that the likelihood function of $\boldsymbol{\theta}\overset{_\Delta}{=}[\hspace*{0.2mm}\boldsymbol{\alpha}_1^{\intercal}\;\dots\;\boldsymbol{\alpha}_n^{\intercal}\;\boldsymbol{\beta}_1^{\intercal}\;\dots\;\boldsymbol{\beta}_{n-1}^{\intercal}\hspace*{0.2mm}]^{\intercal}$ is proportional to
\begin{equation}
\label{likelihoodgeneralizedeq}
L(\boldsymbol{\theta}|\mathbf{Y}) = {\textstyle\prod_{\ell=1}^{L}}
\frac{e^{\sum_{i=1}^{n}\boldsymbol{\alpha}_i^{\intercal}\mathbf{x}_{\ell}^{}d_{i\hspace*{-0.2mm},\ell}^{\alpha}+\boldsymbol{\beta}_i^{\intercal}\mathbf{x}_{\ell}^{}d_{i\hspace*{-0.2mm},\ell}^{\beta}}}{\prod_{i,j=1}^n (1+e^{\boldsymbol{\alpha}_i^{\intercal}\mathbf{x}_{\ell}^{}+\boldsymbol{\beta}_j^{\intercal}\mathbf{x}_{\ell}^{}})^{N_{ij\hspace*{-0.2mm},\ell}}}
\end{equation}
where the outdegree and indegree of node $i$ in random graph $\ell$ have been defined as $d_{i\hspace*{-0.2mm},\ell}^{\alpha}\overset{_\Delta}{=}\sum_{j=1}^{n}Y_{ij\hspace*{-0.2mm},\ell}$ and $d_{i\hspace*{-0.2mm},\ell}^{\beta}\overset{_\Delta}{=}\sum_{j=1}^{n}Y_{ji\hspace*{-0.2mm},\ell}$, respectively. In analogy with the derivation of \eqref{deriveq}, the ML estimates $\hat{\boldsymbol{\theta}}\overset{_\Delta}{=}[\hspace*{0.2mm}\hat{\boldsymbol{\alpha}}_1^{\intercal}\;\dots\;\hat{\boldsymbol{\alpha}}_n^{\intercal}\;\hat{\boldsymbol{\beta}}_1^{\intercal}\;\dots\;\hat{\boldsymbol{\beta}}_{n-1}^{\intercal}\hspace*{0.2mm}]^{\intercal}$ must satisfy $\partial L(\hat{\boldsymbol{\theta}}|\mathbf{Y})/\partial\alpha_{i,k}=0$ for $i=1,\dots,n$ and $k=1,\dots,K$, and $\partial L(\hat{\boldsymbol{\theta}}|\mathbf{Y})/\partial\beta_{i,k}=0$ for $i=1,\dots,n-1$ and $k=1,\dots,K$. This further means that
\begin{subequations}
\label{deriveq2}
\begin{align}
\begin{split}
\sum_{\ell=1}^{L}x_{\ell,k}^{}d_{i\hspace*{-0.2mm},\ell}^{\alpha}=\sum_{\ell=1}^{L}\sum_{j=1}^{n}x_{\ell,k}^{}N_{ij\hspace*{-0.2mm},\ell}^{}\frac{e^{\hat{\boldsymbol{\alpha}}_i^{\intercal}\mathbf{x}_{\ell}^{}+\hat{\boldsymbol{\beta}}_j^{\intercal}\mathbf{x}_{\ell}^{}}}{1+e^{\hat{\boldsymbol{\alpha}}_i^{\intercal}\mathbf{x}_{\ell}^{}+\hat{\boldsymbol{\beta}}_j^{\intercal}\mathbf{x}_{\ell}^{}}},
\end{split}
\intertext{for $i=1,\dots,n$ and $k=1,\dots,K$, where we have used that $\mathbf{x}_{\ell}\overset{_\Delta}{=}[\hspace*{0.2mm}x_{\ell,1}\;\dots\;x_{\ell,K}\hspace*{0.2mm}]^{\intercal}$ and $\hat{\boldsymbol{\beta}}_n\overset{_\Delta}{=}\mathbf{0}_{K\hspace*{-0.2mm},1}$, while}
\begin{split}
\sum_{\ell=1}^{L}x_{\ell,k}^{}d_{i\hspace*{-0.2mm},\ell}^{\beta} =\sum_{\ell=1}^{L}\sum_{j=1}^{n}x_{\ell,k}^{}N_{ji\hspace*{-0.2mm},\ell}\frac{e^{\hat{\boldsymbol{\alpha}}_j^{\intercal}\mathbf{x}_{\ell}^{}+\hat{\boldsymbol{\beta}}_i^{\intercal}\mathbf{x}_{\ell}^{}}}{1+e^{\hat{\boldsymbol{\alpha}}_j^{\intercal}\mathbf{x}_{\ell}^{}+\hat{\boldsymbol{\beta}}_i^{\intercal}\mathbf{x}_{\ell}^{}}},
\end{split}
\end{align}
\end{subequations}
for $i=1,\dots,n-1$ and $k=1,\dots,K$.

We now introduce the function $\boldsymbol{\psi}:\mathbb{R}^{(2n-1)\hspace*{-0.2mm}K}\times\mathbb{R}^{(2n-1)\hspace*{-0.2mm}K}\rightarrow\mathbb{R}^{(2n-1)\hspace*{-0.2mm}K}$, whose outputs are defined according to
\begin{subequations}
\label{psieq}
\begin{align}
\begin{split}
&\hspace*{4.5mm}\psi_{K(i-1)+k}(\mathbf{z},\boldsymbol{\gamma})\\
&\overset{_\Delta}{=}\sum_{\ell=1}^{L}x_{\ell,k}^{}\Bigg(\hspace*{-0.4mm}d_{i\hspace*{-0.2mm},\ell}^{\alpha}-\hspace*{-0.2mm}\sum_{j=1}^{n}\hspace*{-0.2mm}N_{ij\hspace*{-0.2mm},\ell}\frac{e^{\mathbf{z}_i^{\intercal}\mathbf{x}_{\ell}^{}+x_{\ell,k}^{}(\gamma_{i,k}^{}-z_{i,k}^{})}}{e^{-\mathbf{z}_{j+n}^{\intercal}\mathbf{x}_{\ell}^{}}+e^{\mathbf{z}_i^{\intercal}\mathbf{x}_{\ell}^{}}}\Bigg)\hspace*{-0.2mm},\\
\end{split}
\intertext{for $i=1,\dots,n$ and $k=1,\dots,K$, and}
&\hspace*{4.5mm}\psi_{K(i+n-1)+k}(\mathbf{z},\boldsymbol{\gamma})\\
&\overset{_\Delta}{=}\sum_{\ell=1}^{L}x_{\ell,k}\Bigg(\hspace*{-0.4mm}d_{i\hspace*{-0.2mm},\ell}^{\beta}-\hspace*{-0.2mm}\sum_{j=1}^{n}\hspace*{-0.2mm}N_{ji\hspace*{-0.2mm},\ell}\frac{e^{\mathbf{z}_{i+n}^{\intercal}\mathbf{x}_{\ell}^{}+x_{\ell,k}^{}(\gamma_{i+n,k}^{}-z_{i+n,k}^{})}}{e^{-\mathbf{z}_{j}^{\intercal}\mathbf{x}_{\ell}^{}}+e^{\mathbf{z}_{i+n}^{\intercal}\mathbf{x}_{\ell}^{}}}\Bigg),\nonumber
\end{align}
\end{subequations}
for $i=1,\dots,n-1$ and $k=1,\dots,K$. Here, we have used that
$\mathbf{z}\overset{_\Delta}{=}[\hspace*{0.2mm}\mathbf{z}_1^{\intercal}\;\dots\;\mathbf{z}_{2n-1}^{\intercal}\hspace*{0.2mm}]^{\intercal}$, $\mathbf{z}_{2n}\overset{_\Delta}{=}\mathbf{0}_{K\hspace*{-0.2mm},1}$, and $\boldsymbol{\gamma}\overset{_\Delta}{=}[\hspace*{0.2mm}\boldsymbol{\gamma}_1^{\intercal}\;\dots\;\boldsymbol{\gamma}_{2n-1}^{\intercal}\hspace*{0.2mm}]^{\intercal}$, while $\mathbf{z}_i\overset{_\Delta}{=}[\hspace*{0.2mm}z_{i,1}\;\dots\;z_{i,K}\hspace*{0.2mm}]^{\intercal}$ and $\boldsymbol{\gamma}_i\overset{_\Delta}{=}[\hspace*{0.2mm}\gamma_{i,1}\;\dots\;\gamma_{i,K}\hspace*{0.2mm}]^{\intercal}$ for $i=1,\dots,2n-1$. Finally, we define $\boldsymbol{\varphi}:\mathbb{R}^{(2n-1)\hspace*{-0.2mm}K}\rightarrow\mathbb{R}^{(2n-1)\hspace*{-0.2mm}K}$ as
\begin{equation}
\boldsymbol{\varphi}(\mathbf{z})\overset{_\Delta}{=}\{\boldsymbol{\gamma}:\;\boldsymbol{\psi}(\mathbf{z},\boldsymbol{\gamma})=\mathbf{0}_{(2n-1)\hspace*{-0.2mm}K\hspace*{-0.2mm},1}\}.
\end{equation}
In the degenerate case of $L=1$, $K=1$, and $\mathbf{x}_{1}=1$, $\boldsymbol{\varphi}(\mathbf{z})$ reduces to the form given in equation \eqref{phieq}.


Rearranging the terms in \eqref{psieq} and comparing with \eqref{deriveq2} it can be seen that $\hat{\boldsymbol{\theta}}$ is a fixed point of $\boldsymbol{\varphi}(\mathbf{z})$. Hence, we will attempt to find the ML estimates by iterating
\begin{equation}
\label{generalizediterationeq}
\mathbf{z}^{(m+1)}=\boldsymbol{\varphi}(\mathbf{z}^{(m)})
\end{equation}
until convergence, starting from some initial value $\mathbf{z}^{(0)}$. In general, there is no closed-form expression for the elements in $\boldsymbol{\varphi}(\mathbf{z}^{(m)})$. The exceptions are the elements providing updates of eventual intercepts $\alpha_{i,1}$ and $\beta_{i,1}$, i.e., when $x_{\ell,1}=x_{1}$ for $\ell=1,\dots,L$. For these parameters, closed-form expressions similar to those in \eqref{phieq} are always available. For the remaining parameters, we can use that each individual element of $\boldsymbol{\psi}(\mathbf{z},\boldsymbol{\gamma})$ only depends on one matching element in $\boldsymbol{\gamma}$. Therefore, it is easy to numerically find the corresponding elements of $\boldsymbol{\varphi}(\mathbf{z})$ by applying some one-dimensional root-finding algorithm.

\subsection{Implementation}

The initial parameter estimate $\mathbf{z}^{(0)}$ was defined as the zero vector in all three models. The estimates were then iteratively updated as $\mathbf{z}^{(m+1)}=\boldsymbol{\varphi}(\mathbf{z}^{(m)})$ until $\Vert\mathbf{z}^{(m+1)}-\mathbf{z}^{(m)}\Vert<\epsilon$, where the threshold was set to $\epsilon=10^{-4}$. Here, $\Vert\cdot\Vert$ denotes the Euclidean norm.

Several alternative methods, including Newton's method, may be used to find the ML estimates. However, as was hinted in \cite{Holland1981}, a standard implementation of Newton's method will often lead to divergence in low signal-to-noise ratio environments, i.e., where the edge probabilities are close to $0$ or $1$ and the number of observations is small. According to the authors' experience, the fixed-point method presented in this section has more favorable convergence properties. This could also have been expected given the convergence theorems, earlier mentioned in Section \ref{undirbetasec}, that were presented for iteration \eqref{phifncundirected} in \cite{Chatterjee2011} and \cite{Csiszar2011}.


\section{Cram\'er-rao bounds}
\label{CRBsection}

A common way to assess the performance of an estimator is to compare its mean square error (MSE) to the CRB. The CRB provides a lower bound on the MSE of any unbiased estimator, and can be computed directly from the likelihood function. What is more, the CRB can often be used to characterize the estimation problem in terms of its underlying parameters. We write the bound as \cite{Kay1993}
\begin{equation}
\mathrm{Cov}(\hat{\boldsymbol{\theta}})\succeq \mathbf{P}
\end{equation}
where $\mathbf{P}$ is the inverse of the Fisher
information matrix (FIM), and we have used $\mathbf{A}\succeq \mathbf{B}$ to denote
that $\mathbf{A}-\mathbf{B}$ is positive semidefinite. The FIM is defined as
$\boldsymbol{\mathcal{I}}\overset{_\Delta}{=}\mathbb{E}[\hspace*{0.2mm}\mathbf{s}\mathbf{s}^{\intercal}\hspace*{0.2mm}]$ where $\mathbf{s}$ is the gradient of the log-likelihood function (the score function) with respect to $\boldsymbol{\theta}$. In this section, we will derive CRBs for the three models discussed in Section \ref{FrameworkSection}, and use simulations to compare the performance of the ML estimators to the CRBs. Each subsection uses the same notation as in the corresponding subsection of Section \ref{FrameworkSection}. Similarly, the notation for FIMs and inverse FIMs is defined independently in each subsection.

\subsection{The Undirected $\beta$-model}
\label{CRBundirectedsection}

In the undirected $\beta$-model, the element in the $i$th row and $j$th column of the FIM takes the form
\begin{align}
\label{FIMundirected}
\mathcal{I}_{ij}=\begin{cases}
\sum_{q=1}^{n}N_{iq}p_{iq}(1-p_{iq}),&\;i=j \\
N_{ij}p_{ij}(1-p_{ij}),&\;i\neq j.
\end{cases}
\end{align}
Expressed in words, the Fisher information for $\beta_i$ is the sum of variances of the associated edge observations $Y_{i1},\dots,Y_{in}$. Similarly, the Fisher information shared between $\beta_i$ and $\beta_j$ is equal to the variance of $Y_{ij}$. A derivation of \eqref{FIMundirected} is provided in Appendix A.

To demonstrate the effect that the parameters have on the estimation performance, let us consider the special case of $N_{ij}=N$ for all $i\neq j$, $n>2$, $\beta_i=\beta$ for $i=1,\dots,n$, and hence also $p_{ij}=p\overset{_\Delta}{=}e^{2\beta}/(1+e^{2\beta})$ for all $i\neq j$. As shown in Appendix A, the inverse FIM then becomes
\begin{equation}
\label{inverseFIMundirectedeq}
\mathbf{P}=\frac{1}{N\hspace*{-0.2mm}p(1-p)}\frac{1}{(n-2)}\bigg(\mathbf{I}_n-\frac{1}{2(n-1)}\mathbbm{1}_n\bigg).
\end{equation}
Here, $\mathbf{I}_n$ denotes the identity matrix of dimension $n$, while $\mathbbm{1}_n$ denotes the $n\times n$-dimensional matrix with all elements equal to one. Studying the diagonal elements of $\mathbf{P}$, it can be seen that the variance of the individual parameter estimates are bounded from below according to
\begin{align}
\label{scalarCRBundirected}
\begin{split}
\mathrm{Var}(\hat{\beta}_i)&\ge \frac{1}{N\hspace*{-0.2mm}p(1-p)}\frac{2n-3}{2(n-1)(n-2)}\\
&\simeq\frac{1}{N\hspace*{-0.2mm}p(1-p)}\frac{1}{n}
\end{split}
\end{align}
where $\simeq$ denotes asymptotic equality in the limit of $n\rightarrow \infty$. Hence, we see that the CRB is, in the limit of $n\rightarrow \infty$, inversely proportional to both the number of observations $N$ and the number of nodes $n$. Further, the CRB attains its minimum for $p=1\hspace*{-0.2mm}/2$ ($\beta=0$), and approaches infinity both as $p\rightarrow 1$ ($\beta\rightarrow\infty$) and $p\rightarrow 0$ ($\beta\rightarrow-\infty$). Note that we do not assume knowledge of the fact that all parameters $\beta_1,\dots,\beta_n$ have the same value. When assuming that this is known, the model reduces to the Erd\H{o}s-R\'enyi model \cite{Hofstad2015}.

\begin{figure}[t]
\hspace*{-1.5mm}
\vspace*{0.1mm}
\psfragfig[scale=0.6]{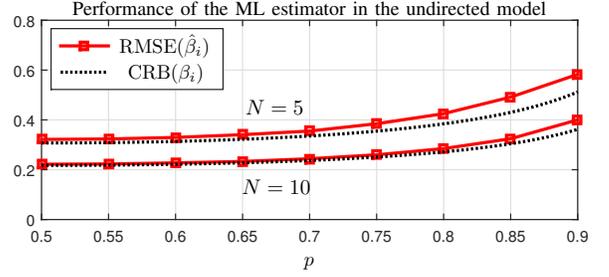}
{
\psfrag{xlabel}[][][0.75]{\raisebox{-3.5mm}{$p$}}
\psfrag{ylabel}[][t][0.75]{}
\psfrag{legendlegendle1}[][][0.75]{\raisebox{1.8mm}{$\mathrm{RMSE}(\hat{\beta}_i)$}}
\psfrag{legendlegendle2}[][][0.75]{\raisebox{1.0mm}{$\mathrm{CRB}(\beta_i)$}}
\psfrag{text1}[][][0.8]{\raisebox{1.3mm}{$N=5$}}
\psfrag{text2}[][][0.8]{\raisebox{1.3mm}{$N=10$}}
\psfrag{title}[][][0.75]{\raisebox{-3.5mm}{Performance of the ML estimator in the undirected model}}
}
\vspace*{-6mm}
\caption{Estimation performance with $n=10$ as dependent on the edge probability $p$.}
\label{simundir}
\vspace*{-1mm}
\end{figure}

Now, continuing with the same the parameter restrictions, Fig. \ref{simundir} shows how the root-mean-square error (RMSE) of the ML estimator compares to the CRB for $N=5$ and $N=10$. Here, the RMSE was computed from $10^4$ simulations\footnote{To avoid non-existence of finite ML estimates, we disregarded any simulation where some node degree was equal to zero or equal to its maximum possible value.} with $n=10$,
and $\mathrm{CRB}(\beta_i)$ is used to denote the square root of the right-hand side of equation \eqref{scalarCRBundirected}. Due to symmetry, we only considered $p\ge 0.5$. As can be seen from Fig. \ref{simundir}, the ML estimator slightly exceeds the CRB when $N=5$, and closely follows the CRB when $N=10$. In both cases, the discrepancy between the RMSE and the CRB increases as $p$ approaches $1$.

\subsection{The Directed $\beta$-model}

As shown in Appendix B, the FIM for the directed $\beta$-model is
\begin{equation}
\label{FIMdirectedeq}
\boldsymbol{\mathcal{I}}=\left[\begin{matrix}
\boldsymbol{\mathcal{I}}_{\alpha} & \boldsymbol{\mathcal{I}}_{\alpha,\hspace*{0.2mm}\beta} \\
\boldsymbol{\mathcal{I}}_{\alpha,\hspace*{0.2mm}\beta}^{\intercal} & \boldsymbol{\mathcal{I}}_{\beta}
\end{matrix}\right]
\end{equation}
where $\boldsymbol{\mathcal{I}}_{\alpha}$ is the diagonal matrix of dimension $n$ whose $i$th diagonal element is $\sum_{j=1}^{n}N_{ij}p_{ij}(1-p_{ij})$, $\boldsymbol{\mathcal{I}}_{\alpha,\hspace*{0.2mm}\beta}$ is the matrix of dimension $n\times n-1$ whose element in the $i$th row and $j$th column is $N_{ij}p_{ij}(1-p_{ij})$, and $\boldsymbol{\mathcal{I}}_{\beta}$ is the diagonal matrix of dimension $n-1$ whose $i$th diagonal element is $\sum_{j=1}^{n}N_{ji}p_{ji}(1-p_{ji})$.

Now, consider the special case when
$N_{ij}=N$ for all $i\neq j$, $n>2$, $\alpha_i=\alpha$ for $i=1,\dots,n$, and $\beta_i=\beta$ for $i=1,\dots,n-1$. Without loss of generality, we assume that $\beta=0$, so that $p_{ij}=p\overset{_\Delta}{=}e^{\alpha}/(1+e^{\alpha})$ for all $i\neq j$. The inverse FIM then becomes
\begin{equation}
\label{inverseFIMdirectedeq}
\mathbf{P}=\frac{1}{N\hspace*{-0.2mm}p(1-p)}\left[\begin{matrix}
\mathbf{P}_{\alpha} & \mathbf{P}_{\alpha,\hspace*{0.2mm}\beta} \\
\mathbf{P}_{\alpha,\hspace*{0.2mm}\beta}^{\intercal} & \mathbf{P}_{\beta}
\end{matrix}\right]
\end{equation}
where
\begin{equation}
\label{Padirected}
\mathbf{P}_{\alpha}\overset{_\Delta}{=}\left[\begin{matrix}
\mathbf{P}_{\alpha,1} & \mathbf{P}_{\alpha,2} \\ \mathbf{P}_{\alpha,2}^{\intercal} & \mathbf{P}_{\alpha,3}
\end{matrix}\right]
\end{equation}
with
\begin{subequations}
\label{Padirectedsub}
\begin{align}
\mathbf{P}_{\alpha,1}&\overset{_\Delta}{=}\frac{n-1}{n(n-2)}\bigg(\hspace*{-0.2mm}\mathbf{I}_{n-1}+\frac{n^2-3n+1}{(n-1)^2}\mathbbm{1}_{n-1}\bigg), \\
\mathbf{P}_{\alpha,2}&\overset{_\Delta}{=}\frac{1}{n-1}\mathbbm{1}_{n-1,1}, \\
\mathbf{P}_{\alpha,3}&\overset{_\Delta}{=}\frac{2n-3}{(n-1)(n-2)},
\end{align}
\end{subequations}
and
\begin{subequations}
\label{PabPbdirected}
\begin{align}
\mathbf{P}_{\alpha,\hspace*{0.2mm}\beta}&\overset{_\Delta}{=}-\frac{1}{n(n-2)}\hspace*{-0.5mm}\left[\begin{matrix}
(n-1)\hspace*{0.2mm}\mathbbm{1}_{n-1}-\mathbf{I}_{n-1}\\
n\hspace*{0.2mm}\mathbbm{1}_{1,n-1}
\end{matrix}\right]\hspace*{-0.8mm},\\
\mathbf{P}_{\beta}&\overset{_\Delta}{=}\frac{n-1}{n(n-2)}(\hspace*{0.2mm}\mathbf{I}_{n-1}+\mathbbm{1}_{n-1}).
\end{align}
\end{subequations}
Here, we have used $\mathbbm{1}_{n,m}$ to denote the $n\times m$ dimensional matrix with all elements equal to one. Further, studying the diagonal elements of $\mathbf{P}$ we arrive at
\begin{subequations}
\label{scalarCRBdirected}
\begin{align}
\label{scalarCRBdirectedalphai}
\mathrm{Var}(\hat{\alpha}_i)&\ge \frac{1}{N\hspace*{-0.2mm}p(1-p)}\frac{2n-1}{n(n-1)},\\
\intertext{for $i=1,\dots,n-1$,}
\mathrm{Var}(\hat{\alpha}_n)&\ge \frac{1}{N\hspace*{-0.2mm}p(1-p)}\frac{2n-3}{(n-1)(n-2)},\\
\shortintertext{and}
\mathrm{Var}(\hat{\beta}_i)&\ge \frac{1}{N\hspace*{-0.2mm}p(1-p)}\frac{2(n-1)}{n(n-2)},
\end{align}
\end{subequations}
for $i=1,\dots,n-1$. Just as for the undirected model, the CRB is, in the limit of $n\rightarrow \infty$, inversely proportional to both $N$ and $n$, while reaching its minimum for $p=1\hspace*{-0.2mm}/2$ ($\beta=0$). Noting that the CRBs in \eqref{scalarCRBdirected} are asymptotically equivalent to $2/(N\hspace*{-0.2mm}p(1-p)n\hspace*{-0.2mm})$ and comparing with \eqref{scalarCRBundirected}, the asymptotic CRBs on the variance can be seen to be twice as large in the directed $\beta$-model as in the undirected $\beta$-model (assuming the same $N$, $n$, and $p$).

\begin{figure}[t]
\hspace*{-1.5mm}
\vspace*{0.1mm}
\psfragfig[scale=0.6]{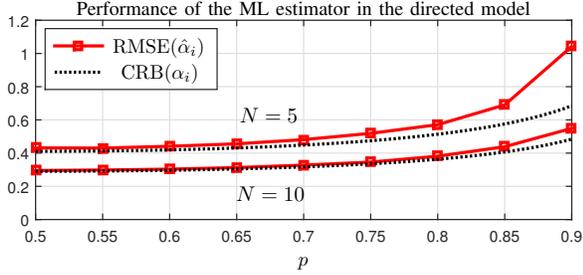}
{
\psfrag{xlabel}[][][0.75]{\raisebox{-3.5mm}{$p$}}
\psfrag{ylabel}[][t][0.75]{}
\psfrag{title}[][][0.75]{\raisebox{-3.5mm}{Performance of the ML estimator in the directed model}}
\psfrag{legendlegendle1}[][][0.75]{\raisebox{1.0mm}{$\mathrm{RMSE}(\hat{\alpha}_i)$}}
\psfrag{legendlegendle2}[][][0.75]{\raisebox{1.0mm}{$\mathrm{CRB}(\alpha_i)$}}
\psfrag{text1}[][][0.8]{\raisebox{1.3mm}{$N=5$}}
\psfrag{text2}[][][0.8]{\raisebox{1.3mm}{$N=10$}}
}
\vspace*{-6mm}
\caption{Estimation performance with $n=10$ and $i\neq n$ as dependent on the edge probability $p$.}
\label{simdir}
\vspace*{-1mm}
\end{figure}

Given the considered parameter restrictions, Fig. \ref{simdir} shows how the ML estimator of $\alpha_i$ with $i\neq n$ compares to the CRB in \eqref{scalarCRBdirectedalphai} for $N=5$ and $N=10$. The RMSE was computed from $10^4$ simulations and the number of nodes was $n=10$. Overall, the estimator displays the same qualitative behavior as for the undirected model in Fig. \ref{simundir}.

\subsection{The Generalized $\beta$-model}
\label{CRBgeneralizedsection}

For the generalized $\beta$-model, we consider the special case when $N_{ij\hspace*{-0.2mm},\ell}=N$ for all $i\neq j$ and $\ell=1,\dots,L$, $n>2$, $\boldsymbol{\alpha}_i=\boldsymbol{\alpha}$ for $i=1,\dots,n$, and $\boldsymbol{\beta}_i=\boldsymbol{\beta}$ for $i=1,\dots,n-1$. Without loss of generality, we assume that $\boldsymbol{\beta}=\mathbf{0}_{K\hspace*{-0.2mm},1}$. Further, this also means that $p_{ij}(\mathbf{x})=p(\mathbf{x})\overset{_\Delta}{=}e^{\boldsymbol{\alpha}^{\intercal}\mathbf{x}}/(1+e^{\boldsymbol{\alpha}^{\intercal}\mathbf{x}})$ for all $i\neq j$. Using $\otimes$ to denote the Kronecker product, the inverse FIM becomes
\begin{equation}
\label{inverseFIMgeneralizedeq}
\mathbf{P}=\frac{1}{N}\left[\begin{matrix}
\mathbf{P}_{\alpha} & \mathbf{P}_{\alpha,\hspace*{0.2mm}\beta} \\
\mathbf{P}_{\alpha,\hspace*{0.2mm}\beta}^{\intercal} & \mathbf{P}_{\beta}
\end{matrix}\right]\otimes \boldsymbol{\mathcal{I}}_x^{-1}
\end{equation}
where $\boldsymbol{\mathcal{I}}_x\overset{_\Delta}{=}\textstyle\sum_{\ell=1}^{L}p(\mathbf{x}_\ell)(1-p(\mathbf{x}_\ell))\mathbf{x}_{\ell}^{}\mathbf{x}_{\ell}^{\intercal}$, while $\mathbf{P}_{\alpha}$, $\mathbf{P}_{\alpha,\hspace*{0.2mm}\beta}$, and $\mathbf{P}_{\beta}$ are as defined in equations \eqref{Padirected}, \eqref{Padirectedsub}, and \eqref{PabPbdirected}. Hence, the CRBs for the regression coefficients pertaining to a given node are subject to the same dependence on $n$ as the regression coefficients pertaining to the corresponding node in the special case of the directed $\beta$-model considered in the preceding subsection. (As a side note, we remark that the analogous statement can be made regarding the CRB for the undirected model presented in Section \ref{CRBundirectedsection} and the CRB for the undirected generalized model, i.e., the model where $\alpha_{i,k}=\beta_{i,k}$ for $i=1,\dots,n$ and $k=1,\dots,K$, while discarding the assumption $\boldsymbol{\beta}_{n}=\mathbf{0}_{K\hspace*{-0.2mm},1}$.) The CRB for the different regression coefficients associated with a given node are then weighted based on
$\boldsymbol{\mathcal{I}}_x^{-1}$. Since the rank of $\boldsymbol{\mathcal{I}}_x$ is $\mathrm{rank}(\boldsymbol{\mathcal{I}}_x)\le\sum_{\ell=1}^{L}\mathrm{rank}\big(p(\mathbf{x}_{\ell})(1-p(\mathbf{x}_{\ell}))\mathbf{x}_{\ell}^{}\mathbf{x}_{\ell}^{\intercal}\big)\le L$, we must have $L\ge K$ for $\boldsymbol{\mathcal{I}}_x$ to be invertible (this condition was expected since the dimension of the observations $\{\{d_{i\hspace*{-0.2mm},\ell}^{\alpha},d_{i\hspace*{-0.2mm},\ell}^{\beta}\}_{i=1}^n\}_{\ell=1}^L$ is $2nL$ and the dimension of $\boldsymbol{\theta}$ is $2nK$). Appendix C first presents the FIM in the general case and then derives the inverse FIM in the special case considered here.

\begin{figure}[t]
\hspace*{-1.5mm}
\vspace*{0.1mm}
\psfragfig[scale=0.6]{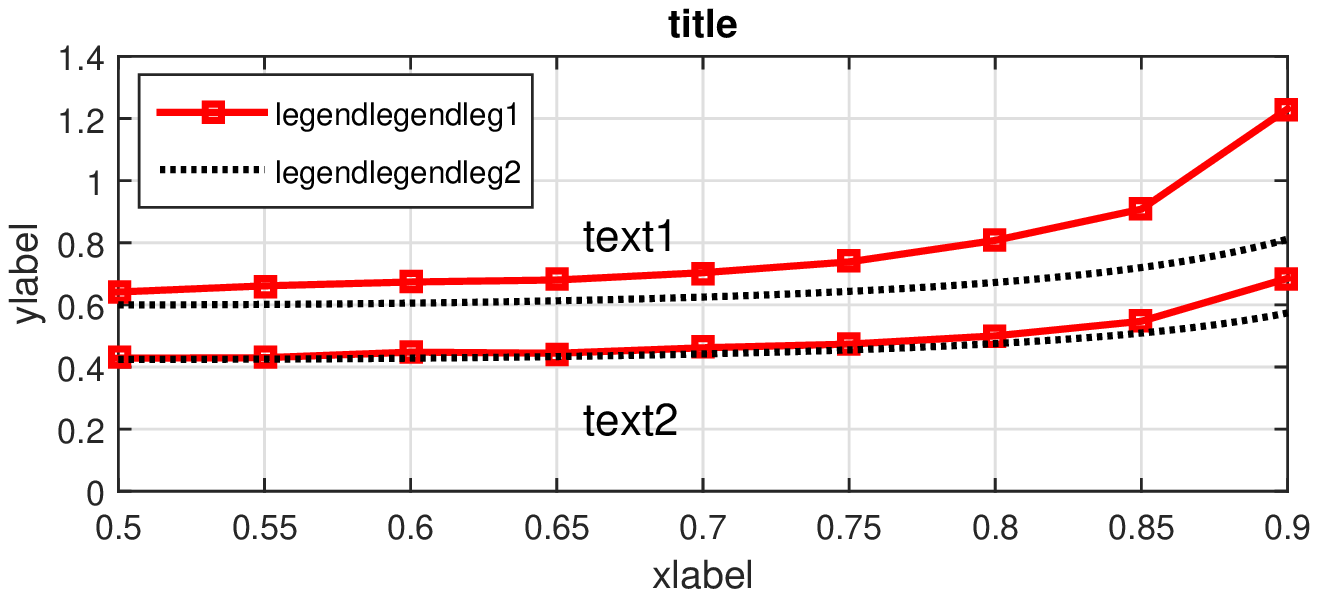}
{
\psfrag{xlabel}[][][0.75]{\raisebox{-3.5mm}{$\bar{p}$}}
\psfrag{ylabel}[][t][0.75]{}
\psfrag{title}[][][0.75]{\raisebox{-3.5mm}{Performance of the ML estimator in the generalized model}}
\psfrag{legendlegendleg1}[][][0.75]{\raisebox{1.0mm}{$\mathrm{RMSE}(\hat{\alpha}_{i1})$}}
\psfrag{legendlegendleg2}[][][0.75]{\raisebox{1.0mm}{$\mathrm{CRB}(\alpha_{i1})$}}
\psfrag{text1}[][][0.8]{\raisebox{1.3mm}{$N=5$}}
\psfrag{text2}[][][0.8]{\raisebox{1.3mm}{$N=10$}}
}
\vspace*{-6mm}
\caption{Estimation performance with $n=5$, $L=2$, $\mathbf{x}_1=[\hspace*{0.2mm}1\;0\hspace*{0.2mm}]^{\intercal}$, $\mathbf{x}_2=[\hspace*{0.2mm}1\;1\hspace*{0.2mm}]^{\intercal}$, $\boldsymbol{\alpha}=\alpha\hspace*{0.3mm}[\hspace*{0.2mm}1\;1\hspace*{0.2mm}]^{\intercal}$, $\boldsymbol{\beta}=\mathbf{0}_{2\hspace*{-0.2mm},1}$, and $i\neq n$, as dependent on the average edge probability $\bar{p}$.}
\label{simgen}
\vspace*{-1mm}
\end{figure}

Now, to illustrate the performance of the ML estimator, we set $n=5$, $L=2$, $\mathbf{x}_1=[\hspace*{0.2mm}1\;0\hspace*{0.2mm}]^{\intercal}$, $\mathbf{x}_2=[\hspace*{0.2mm}1\;1\hspace*{0.2mm}]^{\intercal}$, and
$\boldsymbol{\alpha}=\alpha\hspace*{0.3mm}[\hspace*{0.2mm}1\;1\hspace*{0.2mm}]^{\intercal}$. Fig. \ref{simgen} then shows how the ML estimator of $\alpha_{i1}$ with $i\neq n$ compares to the CRB as $\alpha$ varies. Here, the average edge probability has been defined as $\bar{p}=(p(\mathbf{x}_1)+p(\mathbf{x}_2))/2$. Both the cases $N=5$ and $N=10$ were considered, and the RMSE was computed from $10^4$ simulations. Once again, the RMSE exceeds the CRB for small $N$ and then approaches the CRB as $N$ is modestly increased.

To summarize, this section has presented CRBs for all three studied $\beta$-models. The dependence of the CRBs on the model parameters (number of observations, number of nodes in the graph, and edge probabilities) was shown to be similar across the models. Further, simulations indicated that the ML estimators presented in Section \ref{FrameworkSection} have a RMSE in the vicinity of the CRB already for a comparatively small number of measurements.

\section{Hypothesis testing}

The problem of choosing between two possible hypotheses is often approached by performing a likelihood ratio test (LRT).
This means that the decision is taken based on whether the ratio of the likelihoods of the two hypotheses exceed or fall below some given constant. As stated in the Neyman-Pearson lemma, the LRT is the most powerful of all possible hypothesis tests \cite{Neyman1933}. Put differently, the LRT minimizes the probability of a missed detection (type II error) for a given probability of a false alarm (type I error). When the likelihoods of the hypotheses are dependent on some unknown parameters, it is common to apply the LRT with the parameter values that maximize the respective likelihoods. This is referred to as the generalized likelihood ratio test (GLRT) \cite{Kay1998}. If the two hypotheses can be expressed as ${\cal H}_0:\boldsymbol{\theta}\in\Omega_0$ and ${\cal H}_1:\boldsymbol{\theta}\in\Omega_1$, the generalized likelihood ratio becomes
\begin{equation}
\label{GLR}
\Lambda(\mathbf{Y}) = \frac{\sup_{\boldsymbol{\theta}\in \Omega_0}L(\boldsymbol{\theta}|\mathbf{Y})}{\sup_{\boldsymbol{\theta}\in \Omega_1}L(\boldsymbol{\theta}|\mathbf{Y})}
\end{equation}
and we decide on ${\cal H}_0$ whenever $\Lambda(\mathbf{Y})>\eta$ for some chosen threshold $\eta$.

The threshold $\eta$ is often chosen so that the decision boundary becomes the set of measurements giving some pre-determined $p$-value. However, to be able to compute $p$-values, we need to know the distribution of $\Lambda(\mathbf{Y})$ under ${\cal H}_0$. While the exact distribution is seldom known, approximate $p$-values can often be obtained from Wilks' theorem. Specifically, assume that $\Omega_0$ is a subset of $\Omega_1$ and that $\Omega_1$ has $\kappa$ more free parameters than $\Omega_0$. Then, given certain regularity conditions, Wilks' theorem states that the asymptotic distribution of $-2\log\Lambda_{LR}$ under ${\cal H}_0$, where $\Lambda_{LR}$ is the likelihood ratio, is the $\chi^2$-distribution with $\kappa$ degrees of freedom \cite{Wilks1963}. Analogues to Wilks' theorem, specified for the generalized likelihood ratio, are discussed in \cite{Fan2001}.

Next, we demonstrate how to use the ML estimates presented in Section \ref{FrameworkSection} to perform GLRTs for two simple null hypotheses of practical importance. Goodness-of-fit tests are omitted from the discussion, and we instead refer the reader to earlier studies on ERGMs \cite{Li2013}, the undirected $\beta$-model \cite{Ogawa2013}, and the directed $\beta$-model \cite{Holland1981}, \cite{Petrovic2010}.

\subsection{Significance Tests}

First, we consider the problem of testing whether a specific covariate has any significant impact on the studied graph. To this end, we use the generalized model from Section \ref{generalizedmodelsubsection}, assuming knowledge of $\{\mathbf{x}_{\ell}\}_{\ell=1}^L$ and $\mathbf{Y}=\{\{Y_{ij\hspace*{-0.2mm},\ell}\}_{i,j=1}^n\}_{\ell=1}^L$. Now, to test the significance of the $k$th covariate $x_k$, we formulate the null hypothesis
\begin{equation}
{\cal H}_0: \alpha_{i,k}=\beta_{i,k}=0\;\;\mathrm{for\;}i=1,\dots,n.
\end{equation}
Assuming that the alternative hypothesis ${\cal H}_1$ does not impose any constraints on the parameters $\boldsymbol{\theta}$, it follows that the maximizing parameter in the denominator of \eqref{GLR} is the ML estimate for the generalized $\beta$-model. Similarly, the maximizing parameter in the numerator of \eqref{GLR} is the ML estimate for the generalized $\beta$-model where the employed regression coefficients can be written as $\boldsymbol{\alpha}_i\overset{_\Delta}{=}[\hspace*{0.2mm}\alpha_{i,1}\;\dots\;\alpha_{i,k-1}\;\;\alpha_{i,k+1}\;\dots\;\alpha_{i,K}\hspace*{0.2mm}]^{\intercal}$ and $\boldsymbol{\beta}_i\overset{_\Delta}{=}[\hspace*{0.2mm}\beta_{i,1}\;\dots\;\beta_{i,k-1}\;\;\beta_{i,k+1}\;\dots\;\beta_{i,K}\hspace*{0.2mm}]^{\intercal}$ for $i=1,\dots,n$, while the covariates are $\mathbf{x}_{\ell}\overset{_\Delta}{=}[\hspace*{0.2mm}x_{\ell,1}\;\dots\;x_{\ell,k-1}\;x_{\ell,k+1}\;\dots\;x_{\ell,K}\hspace*{0.2mm}]^{\intercal}$ for $\ell=1,\dots,L$.

\subsection{Testing for Directionality}

Second, we consider the problem of deciding whether some given set of directed network data $\{Y_{ij}\}_{i,j=1}^n$ has originated from a directed or undirected (symmetric) random graph\footnote{A nonparametric test for this problem was developed in \cite{Fagiolo2006}.}. Using the notation of the directed model presented in Section \ref{directedmodelsubsection}, we can write the null hypothesis, indicating that the data originates from an undirected graph, as
\begin{equation}
{\cal H}_0: \alpha_i=\beta_i\;\;\mathrm{for\;}i=1,\dots,n.
\end{equation}
Assuming that the alternative hypothesis ${\cal H}_1$ does not impose any constraints on the parameters $\boldsymbol{\theta}$, it follows that the maximizing parameter in the denominator of \eqref{GLR} is the ML estimate for the directed $\beta$-model. Similarly, the maximizing parameter in the numerator of \eqref{GLR} is the ML estimate for the undirected $\beta$-model, where $Y_{ij}+Y_{ji}$ is considered to be the number of times that an edge between node $i$ and node $j$ is present in $N_{ij}+N_{ji}$ independent measurements. It is trivial to extend the test to the generalized $\beta$-model considered in Section \ref{generalizedmodelsubsection}.

\subsection{Testing for Directionality - Numerical Example}

\begin{figure}[t]
\hspace*{-1.5mm}
\vspace*{0.1mm}
\psfragfig[scale=0.6]{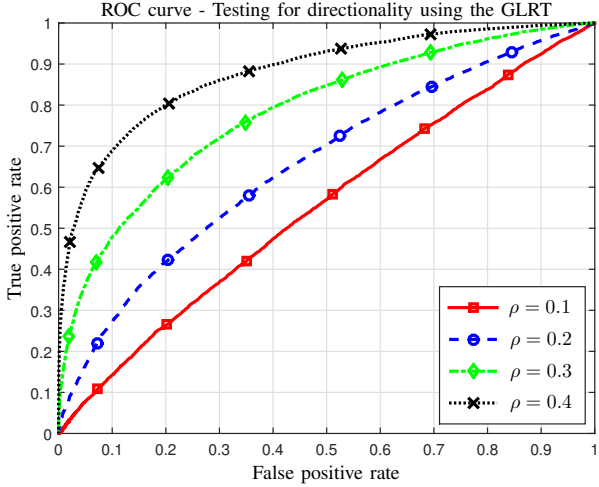}
{
\psfrag{xlabel}[][][0.75]{\raisebox{-3.5mm}{False positive rate}}
\psfrag{ylabel}[][][0.75]{\raisebox{1mm}{True positive rate}}
\psfrag{title}[][][0.75]{\raisebox{-3.8mm}{ROC curve - Testing for directionality using the GLRT}}
\psfrag{legendlege1}[][][0.75]{\raisebox{1.2mm}{$\rho=0.1$}}
\psfrag{legendlege2}[][][0.75]{\raisebox{1.2mm}{$\rho=0.2$}}
\psfrag{legendlege3}[][][0.75]{\raisebox{1.2mm}{$\rho=0.3$}}
\psfrag{legendlege4}[][][0.75]{\raisebox{1.2mm}{$\rho=0.4$}}
}
\vspace*{-6mm}
\caption{ROC curve when testing for directionality in the directed $\beta$-model with $N_{ij}=10$ for all $i\neq j$ and $n=10$.}
\label{rocdirectionality}
\vspace*{-1mm}
\end{figure}

We now study the receiver operating characteristics (ROC) of the GLRT detector, for the problem of testing for directionality, by means of simulations. For the computation of the true positive rate (the probability of correctly rejecting ${\cal H}_0$) we independently simulated $\alpha_i$ for $i=1,\dots,n$ and $\beta_i$ for $i=1,\dots,n-1$ from the uniform distribution over the interval $(-\rho,\rho)$. These parameters were in turn used to simulate $Y_{ij}$ for all $i\neq j$, which then enabled us to compute $\Lambda(\mathbf{Y})$ and, for any given $\eta$, make the decision to reject or accept ${\cal H}_0$. Finally, the true positive rate was obtained as the percentage of rejections. To compute the false positive rate (the probability of incorrectly rejecting ${\cal H}_0$), we simulated $\alpha_i$ for $i=1,\dots,n$ from the uniform distribution over the interval $(-\rho,\rho)$, while letting $\beta_i=\alpha_i$ for $i=1,\dots,n$. After simulating $Y_{ij}$ for all $i\neq j$ (although the distribution of $Y_{ij}$ and $Y_{ji}$ is the same, their simulated values need not be) and computing $\Lambda(\mathbf{Y})$, the false positive rate was obtained as the percentage of rejections of ${\cal H}_0$. Fig. \ref{rocdirectionality} displays the ROC curve for $\rho=0.1,0.2,0.3,0.4$, with $N_{ij}=10$ for all $i\neq j$, $n=10$, when computing each true positive rate and false positive rate from $10^4$ simulations. As expected, the classification performance improves as the simulations of $\alpha_1,\dots,\alpha_n,\beta_1,\dots,\beta_{n-1}$ are drawn from larger intervals, thereby making the sequences $\{\alpha_1,\dots,\alpha_n\}$ and $\{\beta_1,\dots,\beta_n\}$ less alike under the alternative hypothesis ${\cal H}_1$.

\begin{figure}[t]
\hspace*{-1.5mm}
\vspace*{0.1mm}
\psfragfig[scale=0.6]{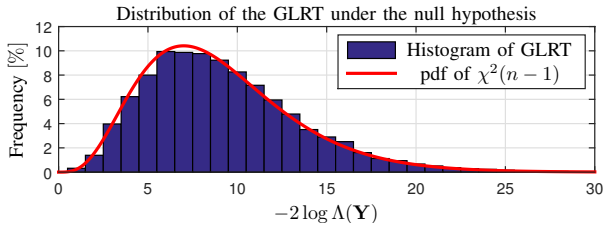}
{
\psfrag{xlabel}[][][0.75]{\raisebox{-4mm}{$-2\log \Lambda (\mathbf{Y})$}}
\psfrag{ylabel}[][][0.75]{\raisebox{2mm}{Frequency$\;[\%]$}}
\psfrag{title}[][][0.75]{\raisebox{-3.8mm}{Distribution of the GLRT under the null hypothesis}}
\psfrag{legendlegendlegendlegend1}[][][0.75]{\raisebox{1.2mm}{Histogram of GLRT}}
\psfrag{legendlegendlegendlegend2}[][][0.75]{\raisebox{1.2mm}{pdf of $\chi^2(n-1)$}}
}
\vspace*{-6mm}
\caption{The distribution of the GLRT under the null hypothesis compared to the $\chi^2$-distribution when testing for symmetry with $N_{ij}=10$ for all $i \neq j$, $n=10$, and $\rho=0.4$.}
\label{wilks}
\vspace*{-1mm}
\end{figure}

Fig. \ref{wilks} displays the histogram of $-2\log\Lambda(\mathbf{Y})$ under ${\cal H}_0$ when testing for directionality with $N_{ij}=10$ for all $i \neq j$, $n=10$, and $\rho=0.4$, using $10^4$ simulations. The histogram is overlaid with the probability density function (pdf) of the $\chi^2$-distribution with $n-1$ degrees of freedom (note that ${\cal H}_0$ is specified by $n$ parameters while ${\cal H}_1$ is specified by $2n-1$ parameters). As can be seen from Fig. \ref{wilks}, the test statistic seems to follow the $\chi^2$-distribution closely, thereby indicating that Wilks' theorem can be of use when performing hypothesis tests under conditions similar to those in the given example.

\section{A case study using social network data}

\begin{figure}[t]
\hspace*{-1.5mm}
\vspace*{0.1mm}
\psfragfig[scale=0.6]{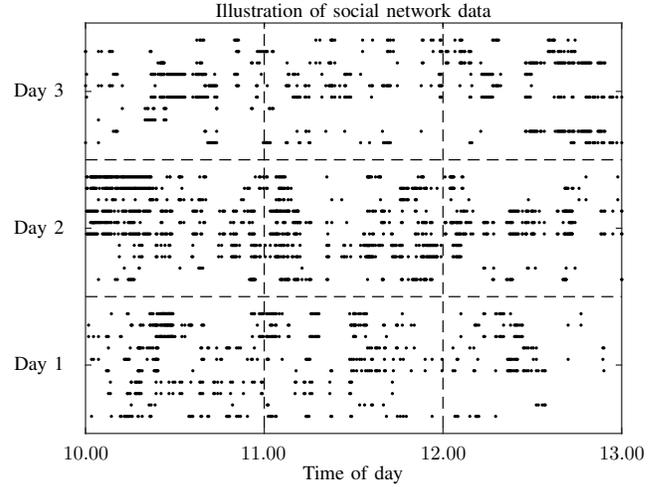}
{
\psfrag{xlabel}[][][0.75]{\raisebox{-3.5mm}{Time of day}}
\psfrag{10}[][][0.75]{\raisebox{-3.5mm}{$10.00$}}
\psfrag{11}[][][0.75]{\raisebox{-3.5mm}{$11.00$}}
\psfrag{12}[][][0.75]{\raisebox{-3.5mm}{$12.00$}}
\psfrag{13}[][][0.75]{\raisebox{-3.5mm}{$13.00$}}
\psfrag{ylabel}[][][0.75]{\raisebox{1mm}{}}
\psfrag{5.5}[r][][0.75]{\raisebox{1mm}{Day 1\hspace*{0.75mm}}}
\psfrag{17.5}[r][][0.75]{\raisebox{1mm}{Day 2\hspace*{0.1mm}}}
\psfrag{29.5}[r][][0.75]{\raisebox{1mm}{Day 3}}
\psfrag{title}[][][0.75]{\raisebox{-3.2mm}{Illustration of social network data}}
}
\vspace*{-6mm}
\caption{Illustration of the studied social network data. Each dot represents a contact made between two individuals.}
\label{dataillustrationFig}
\vspace*{-1mm}
\end{figure}

We now apply the theory presented in the preceding sections to data describing an undirected social network of healthcare workers. Similar data sets have previously been studied using the $\beta$-model in \cite{Holland1981} and \cite{VanDuijn2004}. As noted in \cite{Holland1981}, the $\beta$-model captures the elementary social tendency of differential attraction, while retaining tractability. The data consists of a list of contacts, i.e., events where two individuals were facing each other at a distance less than $1-1.5\,[m]$, and specifies both the time point (a $20$ second interval) at which each contact took place, and the two individuals that were in contact \cite{Vanhems2013}. All in all, contacts among $75$ individuals were recorded over a period of $97$ hours. However, to ensure the existence of all ML estimates, we chose to study the contacts of ten individuals (one administrator, six nurses, and three medical doctors) as registered between $10.00$ and $13.00$ during three consecutive days. Fig. \ref{dataillustrationFig} displays the contacts (made within the smaller group of ten individuals) of a given individual on a specific day along a horizontal line. As can be seen, the studied individuals were all active during the considered time periods. To construct the graph observations, we divided each day into three one-hour periods, with each period generating one observation of each dyad, i.e., each pair of individuals. An edge was considered to be present if and only if there was a contact between the corresponding individuals during the associated time period. Based on the analysis in Section \ref{CRBsection}, a rule of thumb may be to design the model so that roughly half of the observations indicate that a link is present (here, we had $\textstyle\sum_{i,j\hspace*{-0.2mm},\ell} Y_{ij\hspace*{-0.2mm},\ell}/\textstyle\sum_{i,j\hspace*{-0.2mm},\ell} N_{ij\hspace*{-0.2mm},\ell}=0.484$).

Next, we will test how well the studied graphs can be explained by covariates representing the time of day and the day of the week. This is done by incorporating these features as categorical covariates in the generalized $\beta$-model. Specifically, the GLRT is used to test a null hypothesis, under which the distributions of the observations are not dependent on any external covariates, against an alternative hypothesis with one binary covariate indicating whether a given observation is made during a specific day or time of day. This test is then performed for each of the six (the three days and the three time periods) possible covariates. For example, when focusing on the period between $10.00$ and $11.00$, the ML estimates under the alternative hypothesis are computed by using the undirected version of the generalized $\beta$-model with $\mathbf{x}_1=[\hspace*{0.2mm}1\;1\hspace*{0.2mm}]^{\intercal}$ and $\mathbf{x}_2=[\hspace*{0.2mm}1\;0\hspace*{0.2mm}]^{\intercal}$. The contacts displayed in the three leftmost boxes in Fig. \ref{dataillustrationFig} are used to obtain the measurements of the first graph, while the contacts in the remaining boxes provide the measurements of the second graph. Obviously, this means that $N_{ij\hspace*{-0.2mm},\ell}=3\hspace*{0.3mm}\ell$ for all $i\neq j$ and $\ell=1,2$. The ML estimates under the null hypothesis are obtained by using the undirected $\beta$-model and merging all graph data so that $N_{ij}=9$ for all $i\neq j$. The $p$-value of the GLRT can then be estimated in two ways. Firstly, we can study the distribution of the generalized likelihood ratios obtained by simulating measurements from the undirected $\beta$-model. The simulations are made using the ML estimates of the undirected $\beta$-parameters obtained from the real-world data. Secondly, we can employ Wilks' theorem and assume that $-2\log\Lambda(\mathbf{Y})$ is $\chi^2$-distributed with $n=10$ degrees of freedom. Now, Tables \ref{Tab1} and \ref{Tab2} display the simulated and theoretical (as estimated using Wilks' theorem) $p$-values obtained when testing the significance of each of the three time periods (one at a time) and each of the three days (one at a time), respectively. The simulated $p$-values where obtained from $10^4$ simulations. As can be seen, the networks seem to vary considerably with the time of day (two out of three time intervals have a $p$-value smaller than $0.01$), while the day-to-day variations are minor (no day has a $p$-value smaller than $0.05$).

\begin{figure}[t]
\hspace*{-1.5mm}
\vspace*{0.1mm}
\psfragfig[scale=0.6]{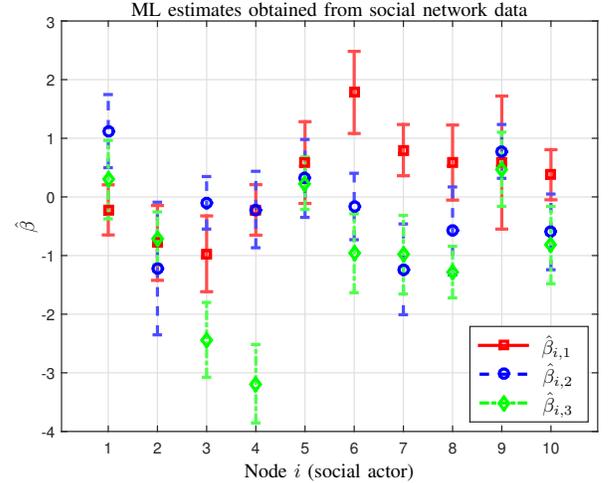}
{
\psfrag{xlabel}[][][0.75]{\raisebox{-3.5mm}{Node $i$ (social actor)}}
\psfrag{ylabel}[][][0.75]{\raisebox{1mm}{$\hat{\beta}$}}
\psfrag{title}[][][0.75]{\raisebox{-3.5mm}{ML estimates obtained from social network data}}
\psfrag{legend1}[][][0.75]{\raisebox{1.2mm}{$\hat{\beta}_{i,1}$}}
\psfrag{legend2}[][][0.75]{\raisebox{1.2mm}{$\hat{\beta}_{i,2}$}}
\psfrag{legend3}[][][0.75]{\raisebox{1.2mm}{$\hat{\beta}_{i,3}$}}
}
\vspace*{-6mm}
\caption{ML estimates obtained when regressing social network data on time of day. The vertical lines represent uncertainty intervals of $\pm$ one standard deviation as estimated from the CRB.}
\label{dataestimates}
\vspace*{-1mm}
\end{figure}

To further investigate how the network depends on the studied time intervals, the network was regressed on all three time-of-day categories simultaneously. Hence, we used the contacts gathered between $10.00$ and $11.00$, between $11.00$ and $12.00$, and between $12.00$ and $13.00$, to obtain measurements of the first, second, and third graph, respectively. The associated covariates where $\mathbf{x}_1=[\hspace*{0.2mm}1\;0\;0\hspace*{0.2mm}]^{\intercal}$, $\mathbf{x}_2=[\hspace*{0.2mm}1\;1\;0\hspace*{0.2mm}]^{\intercal}$, and $\mathbf{x}_3=[\hspace*{0.2mm}1\;0\;1\hspace*{0.2mm}]^{\intercal}$. The resulting estimates are illustrated in Fig. \ref{dataestimates}, where the vertical lines represent uncertainty intervals $[\hat{\beta}_{i,k}-\mathrm{CRB}(\beta_{i,k}),\hat{\beta}_{i,k}+\mathrm{CRB}(\beta_{i,k})]$, and $\mathrm{CRB}(\beta_{i,k})$ denotes the lower bound on the standard deviation of $\hat{\beta}_{i,k}$ as estimated from the undirected version of the FIM presented in \eqref{FIMderivgeneralizedeq} and \eqref{FIMderivgeneralizedeq2}. Here, $\hat{\beta}_{i,1}$ describes the attractiveness of node $i$ in the first time period, while $\hat{\beta}_{i,1}+\hat{\beta}_{i,2}$ and $\hat{\beta}_{i,1}+\hat{\beta}_{i,3}$ describe the attractiveness of node $i$ in the second and third time periods, respectively. Averaging $\hat{\beta}_{i,2}$ and $\hat{\beta}_{i,3}$ over the ten individuals we obtain $-0.19$ and $-0.94$, respectively. In other words, the overall attractiveness is lower in the second time period than in the first, and is lower in the third time period than in the second. This is also what one would have guessed given the distribution of contacts in Fig. \ref{dataillustrationFig} (the total number of observed links in the networks were 81, 71, and 44, in the three respective time periods).

\begin{table}[t]
\caption{$p$-values for social network data. \label{Tab1}}
\centering
\begin{tabular}{l|rrr}
\hline
\hline
& \multicolumn{3}{c}{} \\ [-2.3ex]
&  \multicolumn{3}{c}{Time of day} \\
& & & \\ [-2.5ex]
\cline{2-4}
& & & \\ [-2.3ex]
& \hspace*{-0.5mm}{\footnotesize $10.00\,$\raisebox{0.15mm}{$\mbox{--}$}$\,11.00$}\hspace*{-1mm} &
{\footnotesize $11.00\,$\raisebox{0.15mm}{$\mbox{--}$}$\,12.00$}\hspace*{-1mm} &
{\footnotesize $12.00\,$\raisebox{0.15mm}{$\mbox{--}$}$\,13.00$}\hspace*{-1mm} \\
& & & \\ [-2.7ex]
\hline
& & & \\ [-2.2ex]
\hspace*{-1mm}Simulated $p$-value\hspace*{-1mm}& $0.0051$ & $0.2950$ & $0.0001$ \\ [-0.2ex]
& & & \\ [-2.4ex]
\hspace*{-1mm}Theoretical $p$-value\hspace*{-1mm}& $0.0037$ & $0.2557$ & $5.7\cdot 10^{-6}$ \\ [-0.2ex]
\multicolumn{4}{l}{} \\ [-2.4ex]
\hline \hline
\end{tabular}
\vspace*{-0mm}
\end{table}

\begin{table}[t]
\caption{$p$-values for social network data. \label{Tab2}}
\centering
\begin{tabular}{l|rrr}
\hline
\hline
& \multicolumn{3}{c}{} \\ [-2.3ex]
& \multicolumn{3}{c}{Day} \\
& & & \\ [-2.5ex]
\cline{2-4}
& & & \\ [-2.3ex]
& 1 & 2 & 3 \\
& & & \\ [-2.7ex]
\hline
& & & \\ [-2.2ex]
\hspace*{-1mm}Simulated $p$-value\hspace*{-1mm}& $0.9361$  & $0.1121$ & $0.3125$ \\ [-0.2ex]
& & & \\ [-2.4ex]
\hspace*{-1mm}Theoretical $p$-value\hspace*{-1mm}& $0.9238$  & $0.0943$ & $0.2693$ \\ [-0.2ex]
\multicolumn{4}{l}{} \\ [-2.4ex]
\hline \hline
\end{tabular}
\vspace*{-0mm}
\end{table}

\section{Summary}

In this paper, we generalized the $\beta$-model with the aim of enabling regression of graphs on a given set of covariates. Further, ML estimates and CRBs were derived for the undirected, directed, and generalized $\beta$-model. In the limit where the number of nodes approaches infinity, the CRB on the variance of the $\beta$-parameters was shown to be inversely proportional to both the number of nodes and the number of observations. Moreover, it was demonstrated how to use the ML estimators to perform statistical tests, with simulations indicating that Wilks' theorem can be of use for the computation of $p$-values. A case study where the generalized $\beta$-model was applied to social network data describing contacts among healthcare workers was worked out in detail. In summary, the article presented tractable methods for point estimation, model design, and hypothesis testing in applications studying networks that depend on external covariates. Future research may study the distribution of the ML estimator for the generalized $\beta$-model, as well as alternative implementations and their convergence rates.

%

\section*{Appendix a}

This appendix derives the FIM in \eqref{FIMundirected}, given for the undirected $\beta$-model. Further, we derive the inverse FIM in the special case considered in \eqref{inverseFIMundirectedeq}.

Let us denote the score function as
$\mathbf{s}\overset{_\Delta}{=}[\hspace*{0.2mm}s_1\;\dots\;s_n\hspace*{0.2mm}]^{\intercal}$. As follows directly from the likelihood function in \eqref{likundirectedeq}, it holds that
\begin{equation}
s_i=\sum_{j=1}^{n}Y_{ij}-N_{ij}p_{ij}.
\end{equation}
Further, using \eqref{measdistundirectedeq} and that the measurements from different edges are independent yields
\begin{subequations}
\label{scoreFncResultsundirectedeq}
\begin{equation}
\mathbb{E}[(s_i)^2]=\sum_{j=1}^{n}N_{ij}p_{ij}(1-p_{ij})
\end{equation}
and
\begin{align}
\mathbb{E}[s_is_j]&=\mathbb{E}[(\textstyle\sum_{q=1}^{n}Y_{iq}-N_{iq}p_{iq})(\textstyle\sum_{r=1}^{n}Y_{rj}-N_{rj}p_{rj})]\nonumber\\
&=\mathbb{E}[(Y_{ij}-N_{ij}p_{ij}+ \textstyle\sum_{q\neq j}Y_{iq}-N_{iq}p_{iq})\nonumber\\
&\hspace*{5.5mm}\cdot(Y_{ij}-N_{ij}p_{ij}+\textstyle\sum_{r\neq i}Y_{rj}-N_{rj}p_{rj})]\nonumber\\
&=N_{ij}p_{ij}(1-p_{ij})
\end{align}
\end{subequations}
for any $i\neq j$. The FIM in \eqref{FIMundirected} can now be obtained from \eqref{scoreFncResultsundirectedeq}.

Under the assumptions preceding \eqref{inverseFIMundirectedeq}, it follows that
\begin{equation}
\label{FIMundirectedeq}
\boldsymbol{\mathcal{I}}=N\hspace*{-0.2mm}p(1-p)\big((n-2)\hspace*{0.2mm}\mathbf{I}_n+\mathbbm{1}_n\big).
\end{equation}
Moreover, the Sherman-Morrison formula gives that \cite{Golub1996}
\begin{equation}
\label{ShermanMorrisoneq}
(a\hspace*{0.2mm}\mathbf{I}_n+b\hspace*{0.2mm}\mathbbm{1}_n)^{-1}=\frac{1}{a}\hspace*{0.2mm}\mathbf{I}_n-\frac{b}{a}\frac{1}{a+bn}\mathbbm{1}_n
\end{equation}
for any constants $a\neq 0$ and $b\neq -a/n$. Now, applying the special case of the Sherman-Morrison formula in \eqref{ShermanMorrisoneq} to the FIM in \eqref{FIMundirectedeq} while assuming $n>2$ immediately gives the inverse FIM in \eqref{inverseFIMundirectedeq}.

\section*{Appendix b}

Here, we derive the FIM in \eqref{FIMdirectedeq}, given for the directed $\beta$-model. In addition, we also derive the inverse FIM in the special case considered in \eqref{inverseFIMdirectedeq}.

Let us denote the score function as
$\mathbf{s}\overset{_\Delta}{=}[\hspace*{0.2mm}(\mathbf{s}^{\alpha})^{\intercal}\;(\mathbf{s}^{\beta}\hspace*{0.2mm})^{\intercal}]^{\intercal}$, where $\mathbf{s}^{\alpha}\overset{_\Delta}{=}[\hspace*{0.2mm}s_1^{\alpha}\;\dots\;s_n^{\alpha}\hspace*{0.2mm}]^{\intercal}$ and $\mathbf{s}^{\beta}\overset{_\Delta}{=}[\hspace*{0.2mm}s_1^{\beta}\;\dots\;s_{n-1}^{\beta}\hspace*{0.2mm}]^{\intercal}$. It then follows from \eqref{likelihoodfncdirectedeq} that
\begin{subequations}
\begin{align}
s_i^{\alpha}&=\sum_{j=1}^{n}Y_{ij}-N_{ij}p_{ij},\\
\intertext{for $i=1,\dots,n$, and}
s_i^{\beta}&=\sum_{j=1}^{n}Y_{ji}-N_{ji}p_{ji}.
\end{align}
\end{subequations}
for $i=1,\dots,n-1$. Further, we use \eqref{measdistdirectedeq} and the independence of measurements from different edges to obtain
\begin{subequations}
\label{score1eq}
\begin{align}
\mathbb{E}[(s_i^{\alpha})^2]&=\sum_{j=1}^{n}N_{ij}p_{ij}(1-p_{ij}),\\
\intertext{for $i=1,\dots,n$, and}
\mathbb{E}[(s_i^{\beta})^2]&=\sum_{j=1}^{n}N_{ji}p_{ji}(1-p_{ji}),\\
\mathbb{E}[s_i^{\alpha}s_i^{\beta}]&=0,
\end{align}
\end{subequations}
for $i=1,\dots,n-1$, while
\begin{subequations}
\label{score2eq}
\begin{align}
\mathbb{E}[s_i^{\alpha}s_j^{\alpha}]&=0,\\
\mathbb{E}[s_i^{\beta}s_j^{\beta}]&=0,
\end{align}
and
\begin{align}
\mathbb{E}[s_i^{\alpha}s_j^{\beta}]&=N_{ij}p_{ij}(1-p_{ij}),
\end{align}
\end{subequations}
for any $i\neq j$. The FIM in \eqref{FIMdirectedeq} now follows directly from \eqref{score1eq} and \eqref{score2eq}.

Given the assumptions preceding \eqref{inverseFIMdirectedeq}, it follows that the submatrices in \eqref{FIMdirectedeq} are
\begin{subequations}
\label{FIMsdirected}
\begin{align}
\boldsymbol{\mathcal{I}}_{\alpha}&=N\hspace*{-0.2mm}p(1-p)(n-1)\,\mathbf{I}_{n},\\
\boldsymbol{\mathcal{I}}_{\beta}&=N\hspace*{-0.2mm}p(1-p)(n-1)\,\mathbf{I}_{n-1},\\
\boldsymbol{\mathcal{I}}_{\alpha,\hspace*{0.2mm}\beta}&=N\hspace*{-0.2mm}p(1-p)[\hspace*{0.2mm}
\mathbbm{1}_{n-1}-\mathbf{I}_{n-1} \;\;
\mathbbm{1}_{n-1,1}
\hspace*{0.2mm}]\rule{0pt}{7pt}^{\intercal}.
\end{align}
\end{subequations}
Further, applying the formula for block matrix inversion \cite{Horn1990} and assuming $n>2$, we have that
\begin{align}
\label{inverseFIMeq}
\begin{split}
\mathbf{P}_{\alpha}&=N\hspace*{-0.2mm}p(1-p)(\boldsymbol{\mathcal{I}}_{\alpha}-\boldsymbol{\mathcal{I}}_{\alpha,\hspace*{0.2mm}\beta}^{}\boldsymbol{\mathcal{I}}_{\beta}^{-1}\boldsymbol{\mathcal{I}}_{\alpha,\hspace*{0.2mm}\beta}^{\intercal})^{-1}\\
&=N\hspace*{-0.2mm}p(1-p)\bigg(\hspace*{-0.5mm}\left[\begin{matrix}
\boldsymbol{\mathcal{I}}_{\alpha,1} & \boldsymbol{\mathcal{I}}_{\alpha,2} \\
\boldsymbol{\mathcal{I}}_{\alpha,2}^{\intercal} & \boldsymbol{\mathcal{I}}_{\alpha,3}
\end{matrix}\right]\hspace*{-0.5mm}\bigg)^{-1}
\end{split}
\end{align}
where
\begin{subequations}
\label{FIMmatriceseq}
\begin{align}
\boldsymbol{\mathcal{I}}_{\alpha,1}&\overset{_\Delta}{=}c\hspace*{0.4mm}\big( n(n-2)\hspace*{0.2mm}\mathbf{I}_{n-1}-(n-3)\mathbbm{1}_{n-1}\big),\\
\boldsymbol{\mathcal{I}}_{\alpha,2}&\overset{_\Delta}{=} -c\hspace*{0.4mm}(n-2)\mathbbm{1}_{n-1,1}, \\
\boldsymbol{\mathcal{I}}_{\alpha,3}&\overset{_\Delta}{=}c\hspace*{0.4mm}(n-1)(n-2),
\end{align}
\end{subequations}
with $c\overset{_\Delta}{=}N\hspace*{-0.2mm}p(1-p)/(n-1)$. Here, we have used that
\begin{align}
\begin{split}
&\hspace*{4.5mm}\boldsymbol{\mathcal{I}}_{\alpha,\hspace*{0.2mm}\beta}^{}\boldsymbol{\mathcal{I}}_{\beta}^{-1}\boldsymbol{\mathcal{I}}_{\alpha,\hspace*{0.2mm}\beta}^{\intercal}\\
&=c\hspace*{-0.5mm}\left[\begin{matrix}
\mathbbm{1}_{n-1}-\mathbf{I}_{n-1} \\
\mathbbm{1}_{1,n-1}
\end{matrix}\right]\hspace*{-1mm}
\left[\begin{matrix}
\mathbbm{1}_{n-1}-\mathbf{I}_{n-1} \\
\mathbbm{1}_{1,n-1}
\end{matrix}\right]^{\intercal}\\
&=c\hspace*{-0.5mm}\left[\begin{matrix}
(\mathbbm{1}_{n-1}-\mathbf{I}_{n-1})^2 & (\mathbbm{1}_{n-1}-\mathbf{I}_{n-1})\mathbbm{1}_{n-1,1} \\
\mathbbm{1}_{1,n-1}(\mathbbm{1}_{n-1}-\mathbf{I}_{n-1}) & \mathbbm{1}_{1,n-1}\mathbbm{1}_{n-1,1} \\
\end{matrix}\right]\\
&=c\hspace*{-0.5mm}\left[\begin{matrix}
\mathbf{I}_{n-1}+(n-3)\mathbbm{1}_{n-1} &  (n-2)\mathbbm{1}_{n-1,1} \\
(n-2)\mathbbm{1}_{1,n-1} & n-1 \\
\end{matrix}\right]\hspace*{-0.5mm}.
\end{split}
\end{align}
Using the block matrix inversion formula, it can be seen that
\begin{equation}
\mathbf{P}_{\alpha}\overset{_\Delta}{=}\left[\begin{matrix}
\mathbf{P}_{\alpha,1} & \mathbf{P}_{\alpha,2} \\ \mathbf{P}_{\alpha,2}^{\intercal} & \mathbf{P}_{\alpha,3}
\end{matrix}\right]
\end{equation}
where
\begin{align}
&\hspace*{4.5mm}\mathbf{P}_{\alpha,1}\nonumber\\
&=N\hspace*{-0.2mm}p(1-p)(\boldsymbol{\mathcal{I}}_{\alpha,1}-\boldsymbol{\mathcal{I}}_{\alpha,2}\boldsymbol{\mathcal{I}}_{\alpha,3}^{-1}\boldsymbol{\mathcal{I}}_{\alpha,2}^{\intercal})^{-1}\nonumber\\
&=(n-1)\big(n(n-2)\hspace*{0.2mm}\mathbf{I}_{n-1}-(n-3)\mathbbm{1}_{n-1}\nonumber\\
&\hspace*{4mm}-(n-2)/(n-1)\mathbbm{1}_{n-1}\big)^{-1}\\
&=(n-1)\big(n(n-2)\hspace*{0.2mm}\mathbf{I}_{n-1}-(n^2-3n+1)/(n-1)\mathbbm{1}_{n-1}\big)^{-1}\nonumber\\
&=\frac{n-1}{n(n-2)}\bigg(\hspace*{-0.2mm}\mathbf{I}_{n-1}+\frac{n^2-3n+1}{(n-1)^2}\mathbbm{1}_{n-1}\bigg).\nonumber
\end{align}
Here, the last equality follows from \eqref{ShermanMorrisoneq}. Further, we have
\begin{align}
\mathbf{P}_{\alpha,3}&=N\hspace*{-0.2mm}p(1-p)(\boldsymbol{\mathcal{I}}_{\alpha,3}-\boldsymbol{\mathcal{I}}_{\alpha,2}^{\intercal}\boldsymbol{\mathcal{I}}_{\alpha,1}^{-1}\boldsymbol{\mathcal{I}}_{\alpha,2})^{-1}\nonumber\\
&=\frac{n-1}{n-2}\Big((n-1)-\frac{1}{n}\mathbbm{1}_{1,n-1}\mathbf{I}_{n-1}\mathbbm{1}_{n-1,1}\nonumber\\
&\hspace*{4mm}-\frac{(n-3)}{n(2n-3)}\mathbbm{1}_{1,n-1}\mathbbm{1}_{n-1}\mathbbm{1}_{n-1,1}\Big)^{-1}\\
&=\frac{2n-3}{(n-1)(n-2)}\nonumber
\end{align}
where we have used \eqref{ShermanMorrisoneq} to arrive at
\begin{align}
\begin{split}
\boldsymbol{\mathcal{I}}_{\alpha,1}^{-1}=\frac{1}{cn(n-2)}\Big(\mathbf{I}_{n-1}+\frac{n-3}{2n-3}\mathbbm{1}_{n-1}\Big).
\end{split}
\end{align}
Finally, it holds that
\begin{align}
\mathbf{P}_{\alpha,2}&=-\boldsymbol{\mathcal{I}}_{\alpha,1}^{-1}\boldsymbol{\mathcal{I}}_{\alpha,2}\mathbf{P}_{\alpha,3}\\
&=\frac{1}{n}\Big(\mathbf{I}_{n-1}+\frac{n-3}{2n-3}\mathbbm{1}_{n-1}\Big)\mathbbm{1}_{n-1,1}\mathbf{P}_{\alpha,3}\nonumber\\
&=\frac{1}{n-1}\mathbbm{1}_{n-1,1}\nonumber
\end{align}
which concludes the derivation of $\mathbf{P}_{\alpha}$ as given in equations \eqref{Padirected} and \eqref{Padirectedsub}.

To derive \eqref{PabPbdirected}, we begin by showing that
\begin{align}
\label{Pbetaeq}
\mathbf{P}_{\beta}&=N\hspace*{-0.2mm}p(1-p)(\boldsymbol{\mathcal{I}}_{\beta}-\boldsymbol{\mathcal{I}}_{\alpha,\hspace*{0.2mm}\beta}^{\intercal}\boldsymbol{\mathcal{I}}_{\alpha}^{-1}\boldsymbol{\mathcal{I}}_{\alpha,\hspace*{0.2mm}\beta})^{-1}\nonumber\\
&=\big((n-1)\mathbf{I}_{n-1}-1/(n-1)(\mathbf{I}_{n-1}+(n-2)\mathbbm{1}_{n-1})\big)^{-1}\nonumber\\
&=\frac{n-1}{n-2}\big(n\hspace*{0.2mm}\mathbf{I}_{n-1}-\mathbbm{1}_{n-1}\big)^{-1}\\
&=\frac{n-1}{n(n-2)}(\hspace*{0.2mm}\mathbf{I}_{n-1}+\mathbbm{1}_{n-1})\nonumber
\end{align}
where the second equality uses that
\begin{align}
\begin{split}
\boldsymbol{\mathcal{I}}_{\alpha,\hspace*{0.2mm}\beta}^{\intercal}\boldsymbol{\mathcal{I}}_{\alpha}^{-1}\boldsymbol{\mathcal{I}}_{\alpha,\hspace*{0.2mm}\beta}&=c\hspace*{-0.5mm}\left[\begin{matrix}
\mathbbm{1}_{n-1}-\mathbf{I}_{n-1}\\
\mathbbm{1}_{1,n-1}
\end{matrix}\right]^{\intercal}\hspace*{-0.5mm}
\left[\begin{matrix}
\mathbbm{1}_{n-1}-\mathbf{I}_{n-1} \\
\mathbbm{1}_{1,n-1}
\end{matrix}\right]\\
&=c\hspace*{0.4mm}((\mathbbm{1}_{n-1}-\mathbf{I}_{n-1})^2+\mathbbm{1}_{n-1,1}\mathbbm{1}_{1,n-1})\\
&=c\hspace*{0.4mm}(\hspace*{0.2mm}\mathbf{I}_{n-1}+(n-2)\mathbbm{1}_{n-1})
\end{split}
\end{align}
and the last equality follows from \eqref{ShermanMorrisoneq}. Finally, we have
\begin{align}
\mathbf{P}_{\alpha,\hspace*{0.2mm}\beta}&=-\boldsymbol{\mathcal{I}}_{\alpha}^{-1}\boldsymbol{\mathcal{I}}_{\alpha,\hspace*{0.2mm}\beta}\mathbf{P}_{\beta}\\
&=-\frac{1}{n(n-2)}\hspace*{-0.5mm}\left[\begin{matrix}
\mathbbm{1}_{n-1}-\mathbf{I}_{n-1}\\
\mathbbm{1}_{1,n-1}
\end{matrix}\right]\hspace*{-0.5mm}(\hspace*{0.2mm}\mathbf{I}_{n-1}+\mathbbm{1}_{n-1})\nonumber\\
&=-\frac{1}{n(n-2)}\hspace*{-0.5mm}\left[\begin{matrix}
(n-1)\hspace*{0.2mm}\mathbbm{1}_{n-1}-\mathbf{I}_{n-1}\\
n\hspace*{0.2mm}\mathbbm{1}_{1,n-1}
\end{matrix}\right]\nonumber
\end{align}
and we are done.

\section*{Appendix c}

We now show how to obtain the FIM for the generalized $\beta$-model, and then go on to derive the inverse FIM for the special case considered in \eqref{inverseFIMgeneralizedeq}.

To begin with, let us denote the score function as
$\mathbf{s}\overset{_\Delta}{=}[\hspace*{0.2mm}(\mathbf{s}_1^{\alpha})^{\intercal}\;\dots\;(\mathbf{s}_n^{\alpha})^{\intercal}\;(\mathbf{s}_1^{\beta})^{\intercal}\;\dots\;(\mathbf{s}_{n-1}^{\beta})^{\intercal}\hspace*{0.2mm}]^{\intercal}$ where $\mathbf{s}_i^{\alpha}\overset{_\Delta}{=}[\hspace{0.2mm}s_{i,1}^{\alpha}\;\dots\;s_{i,K}^{\alpha}\hspace{0.2mm}]^{\intercal}$ for $i=1,\dots,n$ and $\mathbf{s}_i^{\beta}\overset{_\Delta}{=}[\hspace{0.2mm}s_{i,1}^{\beta}\;\dots\;s_{i,K}^{\beta}\hspace{0.2mm}]^{\intercal}$ for $i=1,\dots,n-1$. It then follows from \eqref{likelihoodgeneralizedeq} that
\begin{subequations}
\begin{align}
s_{i,k}^{\alpha}&=\sum_{\ell=1}^{L}\sum_{j=1}^n x_{\ell,k}^{}(Y_{ij\hspace*{-0.2mm},\ell}-N_{ij\hspace*{-0.2mm},\ell}\hspace*{0.2mm}p_{ij}(\mathbf{x}_{\ell}))\\
\intertext{for $i=1,\dots,n$ and $k=1,\dots,K$, while}
s_{i,k}^{\beta}&=\sum_{\ell=1}^{L}\sum_{j=1}^n x_{\ell,k}^{}(Y_{ji\hspace*{-0.2mm},\ell}-N_{ji\hspace*{-0.2mm},\ell}\hspace*{0.2mm}p_{ji}(\mathbf{x}_{\ell}))
\end{align}
\end{subequations}
for $i=1,\dots,n-1$ and $k=1,\dots,K$. Further, using \eqref{measdistgeneralizedeq} and the independence of measurements from different edges and different graphs we have that
\begin{subequations}
\label{FIMderivgeneralizedeq}
\begin{align}
\mathbb{E}[s_{i,k}^{\alpha}s_{i,k'}^{\alpha}]
&=\sum_{\ell=1}^{L}\sum_{j=1}^{n}x_{\ell,k}^{}x_{\ell,k'}^{}N_{ij\hspace*{-0.2mm},\ell}\hspace*{0.2mm}p_{ij}(\mathbf{x_{\ell}})(1-p_{ij}(\mathbf{x_{\ell}}))\\
\intertext{for $i=1,\dots,n$, $k=1,\dots,K$, and $k'=1,\dots,K$, while}
\mathbb{E}[s_{i,k}^{\beta}s_{i,k'}^{\beta}]
&=\sum_{\ell=1}^{L}\sum_{j=1}^{n}x_{\ell,k}^{}x_{\ell,k'}^{}N_{ji\hspace*{-0.2mm},\ell}\hspace*{0.2mm}p_{ji}(\mathbf{x_{\ell}})(1-p_{ji}(\mathbf{x_{\ell}})),
\end{align}
and
\begin{equation}
\mathbb{E}[s_{i,k}^{\alpha}s_{i,k'}^{\beta}]=0,
\end{equation}
\end{subequations}
for $i=1,\dots,n-1$, $k=1,\dots,K$, and $k'=1,\dots,K$. Moreover, we have that
\begin{subequations}
\label{FIMderivgeneralizedeq2}
\begin{align}
\mathbb{E}[s_{i,k}^{\alpha}s_{j,k'}^{\alpha}]&=0,\\
\mathbb{E}[s_{i,k}^{\beta}s_{j,k'}^{\beta}]&=0,
\end{align}
and
\begin{align}
\mathbb{E}[s_{i,k}^{\alpha}s_{j,k'}^{\beta}]=\sum_{\ell=1}^{L}x_{\ell,k}^{}x_{\ell,k'}^{}N_{ij\hspace*{-0.2mm},\ell}\hspace*{0.2mm}p_{ij}(\mathbf{x_{\ell}})(1-p_{ij}(\mathbf{x_{\ell}})),
\end{align}
\end{subequations}
for any $i\neq j$, $k=1,\dots,K$, and $k'=1,\dots,K$. The FIM for the generalized $\beta$-model can now be obtained from equations \eqref{FIMderivgeneralizedeq} and \eqref{FIMderivgeneralizedeq2}.

For the special case considered in Section \ref{CRBgeneralizedsection}, it is easily shown that the FIM reduces to
\begin{equation}
\boldsymbol{\mathcal{I}}=\left[\begin{matrix}
\boldsymbol{\mathcal{I}}_{\alpha} & \boldsymbol{\mathcal{I}}_{\alpha,\hspace*{0.2mm}\beta} \\
\boldsymbol{\mathcal{I}}_{\alpha,\hspace*{0.2mm}\beta}^{\intercal} & \boldsymbol{\mathcal{I}}_{\beta}
\end{matrix}\right]\otimes\boldsymbol{\mathcal{I}}_x
\end{equation}
where
\begin{subequations}
\label{FIMsgeneralized}
\begin{align}
\boldsymbol{\mathcal{I}}_{\alpha}&\overset{_\Delta}{=}N\hspace*{-0.2mm}(n-1)\,\mathbf{I}_n, \\
\boldsymbol{\mathcal{I}}_{\beta}&\overset{_\Delta}{=}N\hspace*{-0.2mm}(n-1)\,\mathbf{I}_{n-1}, \\
\boldsymbol{\mathcal{I}}_{\alpha,\hspace*{0.2mm}\beta}&\overset{_\Delta}{=}N[\hspace*{0.2mm} \mathbbm{1}_{n-1}-\mathbf{I}_{n-1} \;
\mathbbm{1}_{n-1,1}\hspace*{0.2mm}]\rule{0pt}{7pt}^{\intercal},
\end{align}
\end{subequations}
and $\boldsymbol{\mathcal{I}}_x\overset{_\Delta}{=}\textstyle\sum_{\ell=1}^{L}p(\mathbf{x}_\ell)(1-p(\mathbf{x}_\ell))\mathbf{x}_{\ell}^{}\mathbf{x}_{\ell}^{\intercal}$. Noting the similarity of equations \eqref{FIMsdirected} and \eqref{FIMsgeneralized}, while using that $(\mathbf{A}\otimes\mathbf{B})^{-1}=\mathbf{A}^{-1}\otimes \mathbf{B}^{-1}$ for any invertible matrices $\mathbf{A}$ and $\mathbf{B}$, the FIM in \eqref{inverseFIMgeneralizedeq} follows immediately when assuming that $\boldsymbol{\mathcal{I}}_x$ is invertible and that $n>2$.

\ifCLASSOPTIONcaptionsoff
  \newpage
\fi



\bibliographystyle{IEEEtran}
\bibliography{refs}

\begin{IEEEbiography}[{\includegraphics[width=1in,height=1.25in,clip,keepaspectratio]{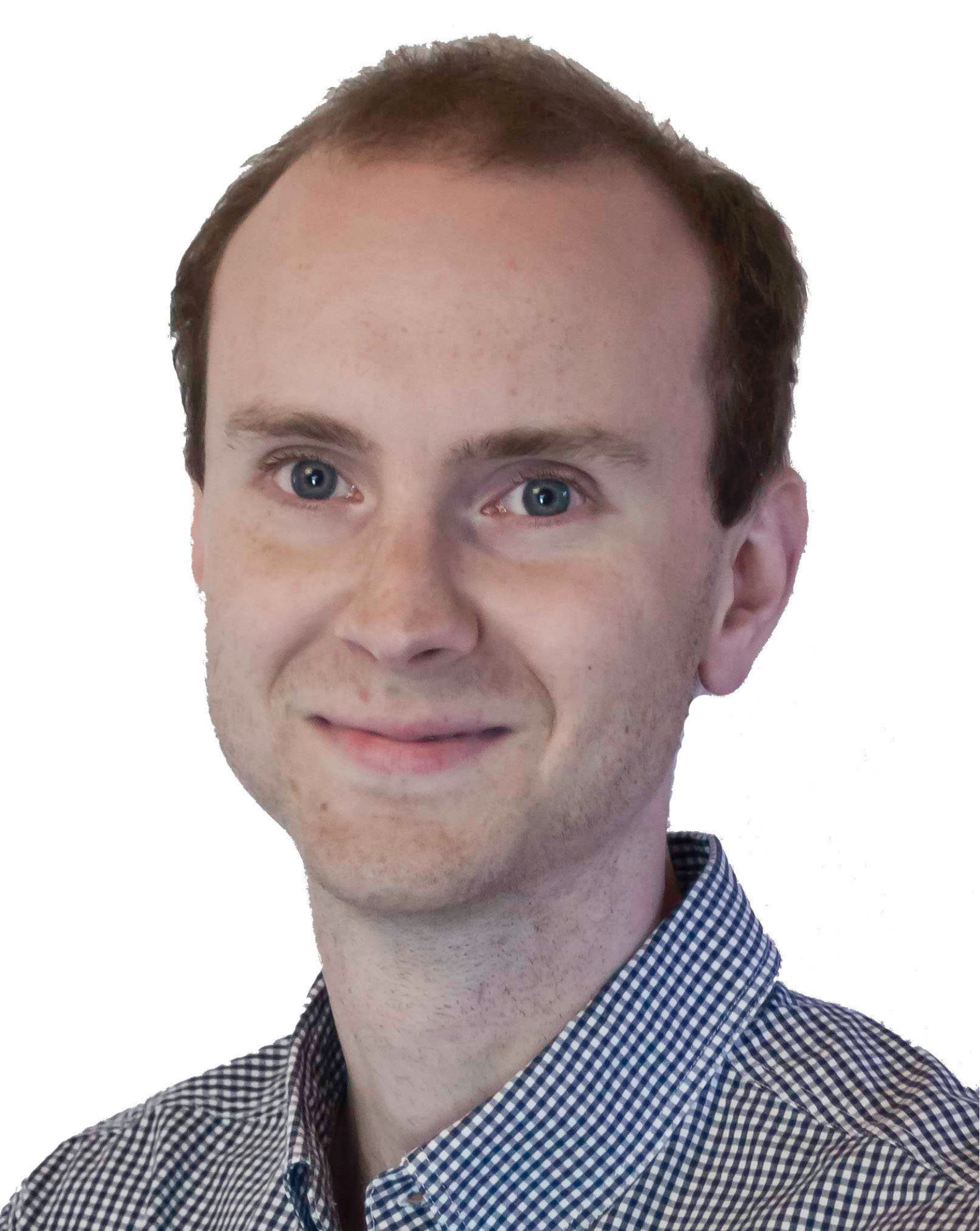}}]
{Johan Wahlström} received his MSc degree in Engineering Physics from the KTH Royal Institute of Technology, Stockholm, Sweden, in 2014. He subsequently joined the Signal Processing Department at KTH, working towards his PhD. His main research topic is insurance telematics. In 2015, he received a scholarship from the Sweden-America foundation and spent six months at Washington University, St. Louis, USA.
\end{IEEEbiography}

\begin{IEEEbiography}[{\includegraphics[width=1in,height=1.25in,clip,keepaspectratio]{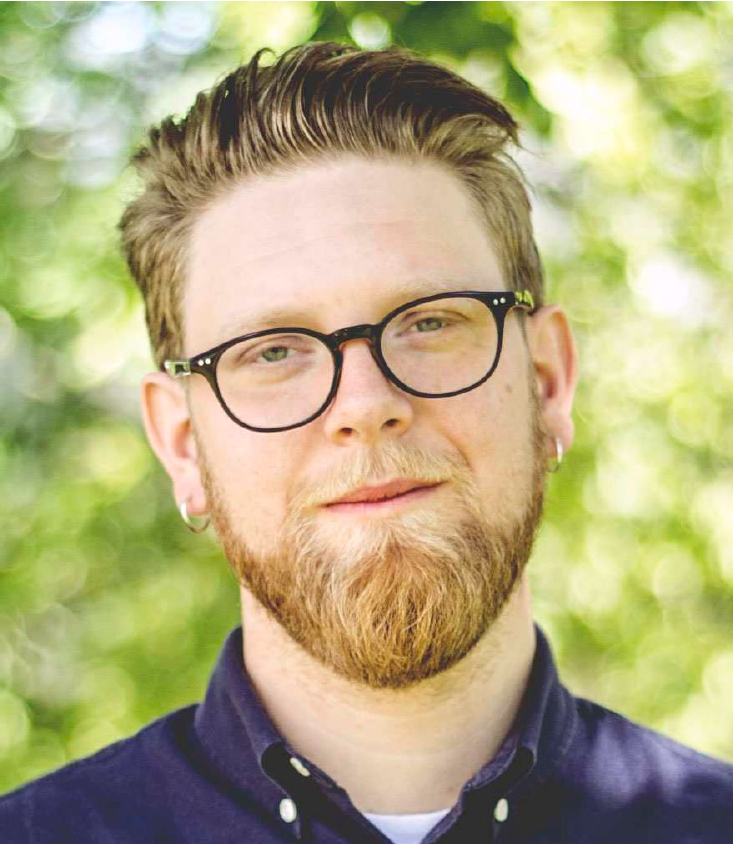}}]
{Isaac Skog}(S'09-M'10) received the BSc and MSc degrees in Electrical Engineering from the KTH Royal Institute of Technology, Stockholm, Sweden, in 2003 and 2005, respectively. In 2010, he received the PhD degree in Signal Processing with a thesis on low-cost navigation systems. In 2009, he spent 5 months at the Mobile Multi-Sensor System research team, University of Calgary, Canada, as a visiting scholar and in 2011 he spent 4 months at the Indian Institute of Science (IISc), Bangalore, India, as a visiting scholar. He is currently a Researcher at KTH coordinating the KTH Insurance Telematics Lab. He was a recipient of a Best Survey Paper Award by the IEEE Intelligent Transportation Systems Society in 2013.
\end{IEEEbiography}

\begin{IEEEbiography}[{\includegraphics[width=1in,height=1.25in,clip,keepaspectratio]{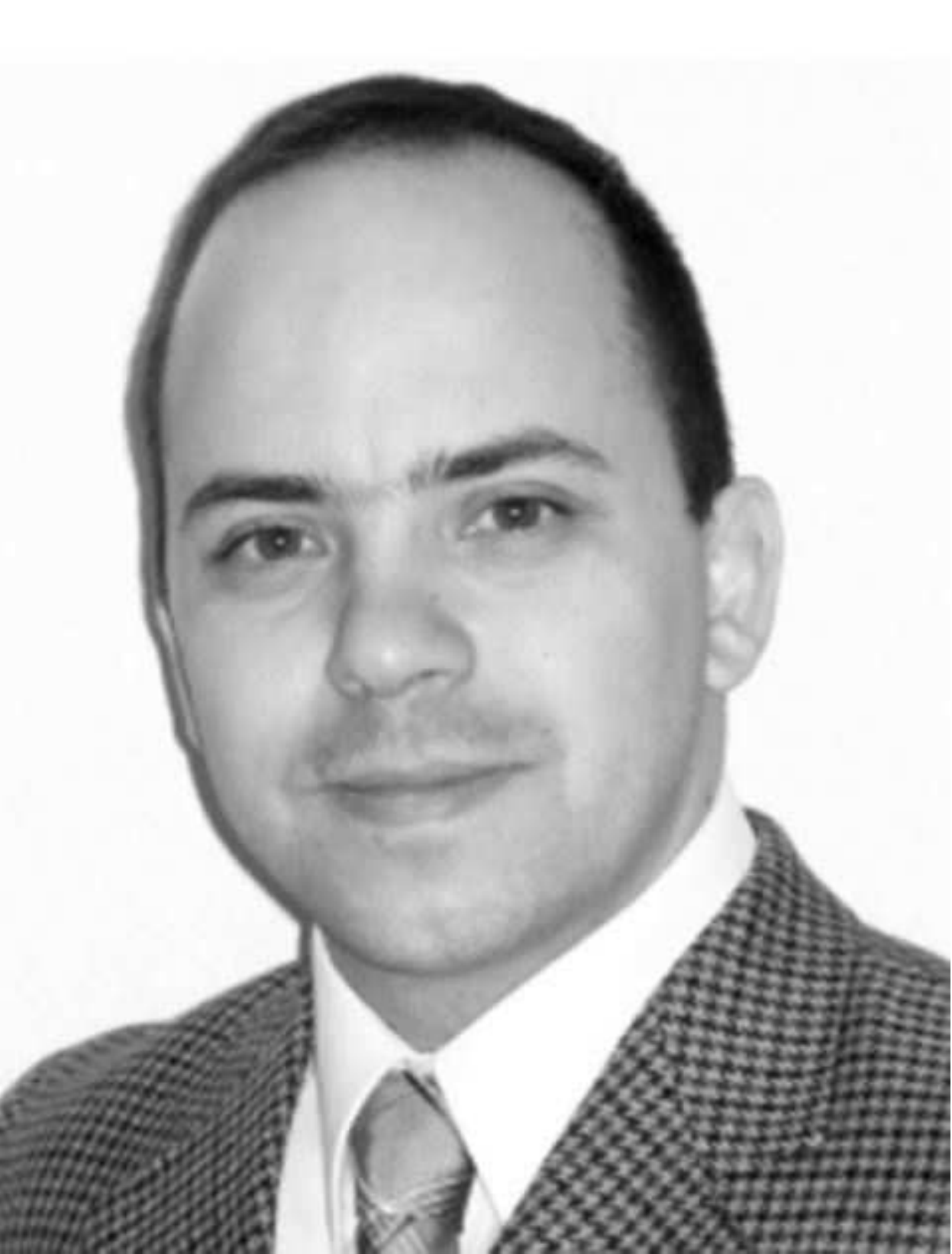}}]
{Patricio S. La Rosa}(S'07-M'09) received the BSc degree in engineering from the Pontifical Catholic
University of Chile (PUC) in 1999, the MSc degree (with maximum distinction) in electrical engineering
from the University of Chile in 2003, and the MSc and the PhD degrees in electrical engineering from
Washington University in St. Louis (WUSTL) in 2010, under the guidance of Prof. A. Nehorai. He was a Research Assistant in the Signal and Image Research Laboratory at the University of Illinois at Chicago, from 2003 to 2005, and in the Center for Sensor Signal and Information Processing at WUSTL, from 2006 to 2010. He has been a Postdoctoral Research Associate at the General Medical Sciences Division, Department of Internal Medicine, Washington University School of Medicine. Currently, he is a Research Data Scientist at Monsanto Company. His research interests are in statistical array signal processing and its applications to medical imaging and biosignal processing, and in biophysical modeling, including nonlinear waveform phenomena in excitable media, bioelectromagnetism, and optical mapping. Dr. La Rosa received the John Paul II Foundation scholarship between the years 1995 and 2000 for undergraduate studies in engineering sciences at PUC.
\end{IEEEbiography}

\begin{IEEEbiography}[{\includegraphics[width=1in,height=1.25in,clip,keepaspectratio]{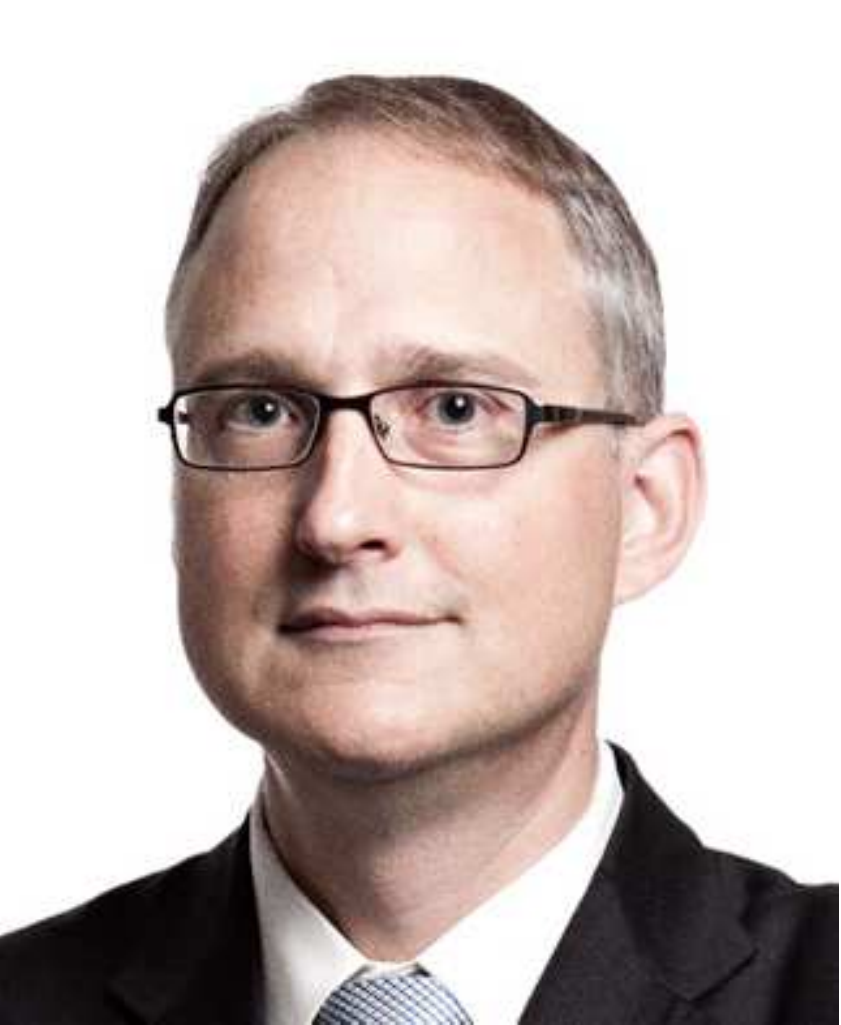}}]
{Peter H\"{a}ndel}(S'88-M'94-SM'98) received a PhD degree from Uppsala University, Uppsala, Sweden, in 1993. From 1987 to 1993, he was with Uppsala University. From 1993 to 1997, he was with Ericsson AB, Kista, Sweden. From 1996 to 1997, he was a Visiting Scholar with the Tampere University of Technology, Tampere, Finland. Since 1997, he has been with the KTH Royal Institute of Technology, Stockholm, Sweden, where he is currently a Professor of Signal Processing and the Head of the Department of Signal Processing. From 2000 to 2006, he held an adjunct position at the Swedish Defence Research Agency. He has been a Guest Professor at the Indian Institute of Science (IISc), Bangalore, India, and at the University of G\"avle, Sweden. He is a co-founder of Movelo AB. Dr. H\"andel has served as an associate editor for the \emph{IEEE Transactions on Signal Processing}. He was a recipient of a Best Survey Paper Award by the IEEE Intelligent Transportation Systems Society in 2013.
\end{IEEEbiography}

\begin{IEEEbiography}[{\includegraphics[width=1in,height=1.25in,clip,keepaspectratio]{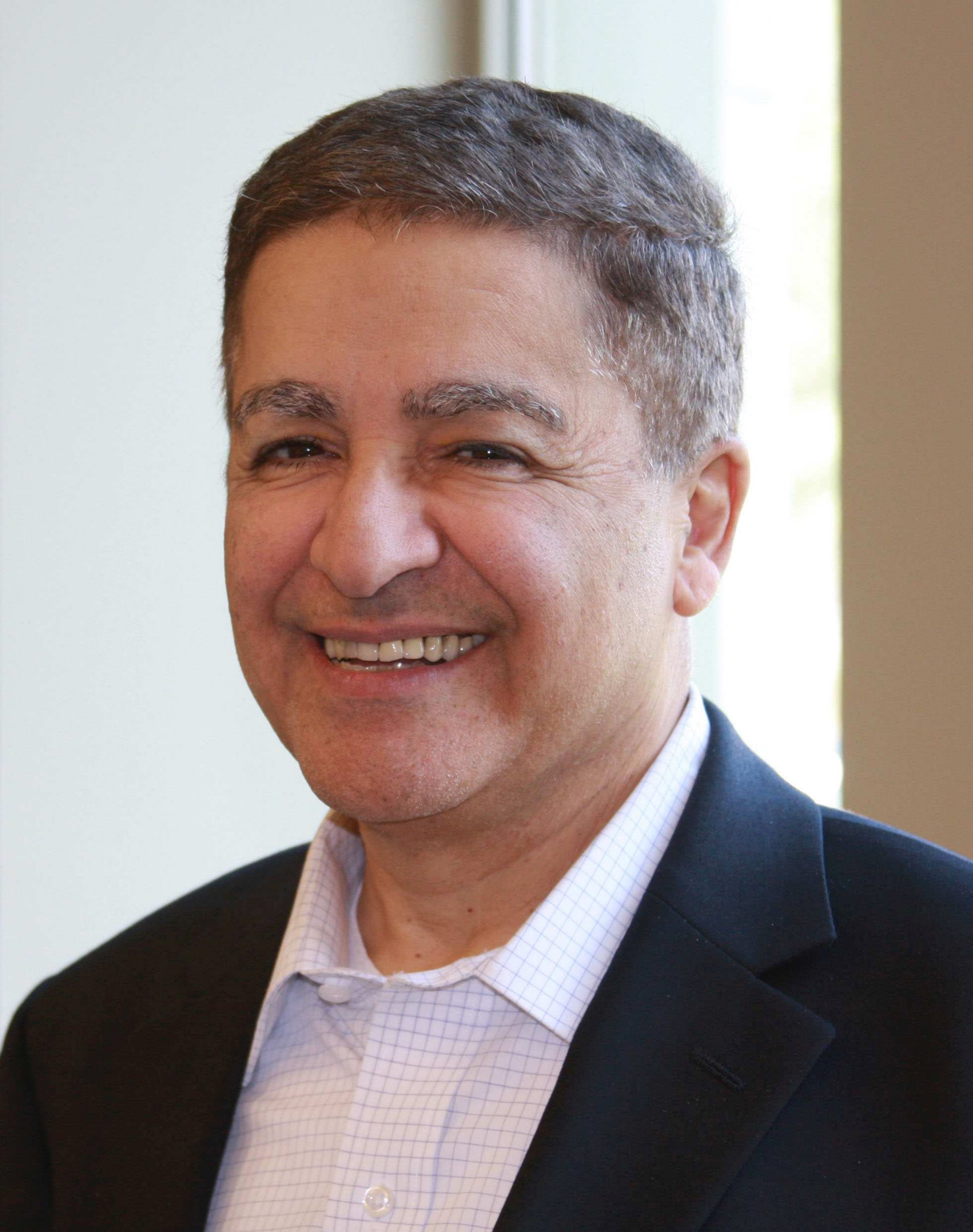}}]
{Arye Nehorai} (S'80-M'83-SM'90-F'94) is the Eugene and Martha Lohman Professor of Electrical Engineering in the Preston M.\ Green Department of Electrical and Systems Engineering (ESE) at Washington University in St.\ Louis (WUSTL). He served as chair of this department from 2006 to 2016. Under his department chair leadership, the undergraduate enrollment has more than tripled in four years and the master's enrollment grew seven-fold in the same time period. He is also Professor in the Department of Biomedical Engineering (by courtesy), Professor in the Division of Biology and Biomedical Studies (DBBS), and Director of the Center for Sensor Signal and Information Processing at WUSTL. Prior to serving at WUSTL, he was a faculty member at Yale University and the University of Illinois at Chicago. He received the BSc and MSc degrees from the Technion, Israel and the PhD from Stanford University, California.

Dr.\ Nehorai served as Editor-in-Chief of the {\em IEEE Transactions on Signal Processing}\/ from 2000 to 2002. From 2003 to 2005 he was the Vice President (Publications) of the IEEE Signal Processing Society (SPS), the Chair of the Publications Board, and a member of the Executive Committee of this Society. He was the founding editor of the special columns on Leadership Reflections in {\em IEEE Signal Processing Magazine}\/ from 2003 to 2006.

Dr.\ Nehorai received the 2006 IEEE SPS Technical Achievement Award and the 2010 IEEE SPS Meritorious Service Award. He was elected Distinguished Lecturer of the IEEE SPS for a term lasting from 2004 to 2005. He received several best paper awards in IEEE journals and conferences. In 2001 he was named University Scholar of the University of Illinois. Dr.\ Nehorai was the Principal Investigator of the Multidisciplinary University Research Initiative (MURI) project titled Adaptive Waveform Diversity for Full Spectral Dominance from 2005 to 2010. He is a Fellow of the IEEE since 1994, Fellow of the Royal Statistical Society since 1996, and Fellow of AAAS since 2012.
\end{IEEEbiography}









\end{document}